\documentclass[seceq,supplement]{ptptex}
\usepackage{bm}		
\usepackage{graphicx}	

\pubinfo{No.~176, 2008}
\title{Electronic Structure of Multilayer Graphene}

\author{Hongki \textsc{Min}\footnote{E-mail: hongki@physics.utexas.edu} 
and Allan H. \textsc{MacDonald}}

\inst{Department of Physics, University of Texas at Austin, Austin, 
Texas 78712, USA}

\abst{We study the electronic structure of multilayer graphene 
using a $\pi$-orbital continuum model with nearest-neighbor intralayer 
and interlayer tunneling.
Using degenerate state perturbation theory, we show that the
low-energy electronic structure of arbitrarily stacked graphene 
multilayers consists of chiral pseudospin doublets with a conserved 
chirality sum. }

\begin{document}
\maketitle

\section{Introduction} 
\label{sec:introduction}
The recent explosion\cite{geim2007a,geim2007b} of research on the electronic properties of single layer and stacked multilayer graphene sheets
has been driven by advances in material preparation methods\cite{novoselov2004,berger2004}, by the unusual\cite{ohta2006,rycerz2007,cheianov2007} electronic properties of these
materials including unusual quantum Hall effects\cite{novoselov2005,zhang2005},
and by hopes that these elegantly tunable systems might be useful electronic materials.

In this paper\cite{min2008}, we study the electronic structure of arbitrarily stacked multilayer graphene using a $\pi$-orbital continuum model with only near-neighbor interactions, analyzing its low-energy spectrum using degenerate state perturbation theory.
Here we focus solely on aligned multilayer graphene without rotational stacking faults\cite{hass2008}.
Interestingly, we find that the low-energy effective theory of multilayer graphene is always described by a set of chiral
pseudospin doublets with a conserved chirality sum.  We discuss implications of this finding for the quantum Hall
effect in multilayer graphene.

\section{$\pi$-orbital continuum model}
\label{sec:pi_orbital_model}

We consider the $\pi$-orbital continuum model for $N$-layer graphene Hamiltonian
which describes bands near the hexagonal corners of the triangular lattice Brillouin zone, 
the $K$ and $K'$ points:
\begin{equation}
{\cal H}=\sum_{\bm p} \Psi_{\bm p}^{\dagger} H({\bm p}) \Psi_{\bm p},
\end{equation}
where $\Psi_{\bm p}=(c_{1,\alpha,{\bm p}},c_{1,\beta,{\bm p}},\cdots,c_{N,\alpha,{\bm p}},c_{N,\beta,{\bm p}})$ and $c_{l,\mu,{\bm p}}$ is an electron annihilation operator
for layer $l=1,\cdots,N$, sublattice $\mu=\alpha,\beta$ and momentum $\bm p$ measured from $K$ or $K'$ point.

The simplest model for a multilayer graphene system allows only  
nearest-neighbor intralayer hopping $t$ and the nearest-neighbor interlayer hopping $t_{\perp}$. 
The in-plane Fermi velocity $v$ is related with $t$ by ${\hbar v \over a}={\sqrt{3}\over 2} t$, 
where $a=2.46$ ${\rm \AA}$ is a lattice constant of monolayer graphene.  Although this model 
is not fully realistic, some aspects of the electronic structure can be understood by 
fully analyzing the properties of this simplified model first and then considering corrections.

\subsection{Stacking diagrams}
\begin{figure}[t]
\begin{center}
\includegraphics[width=0.7\linewidth]{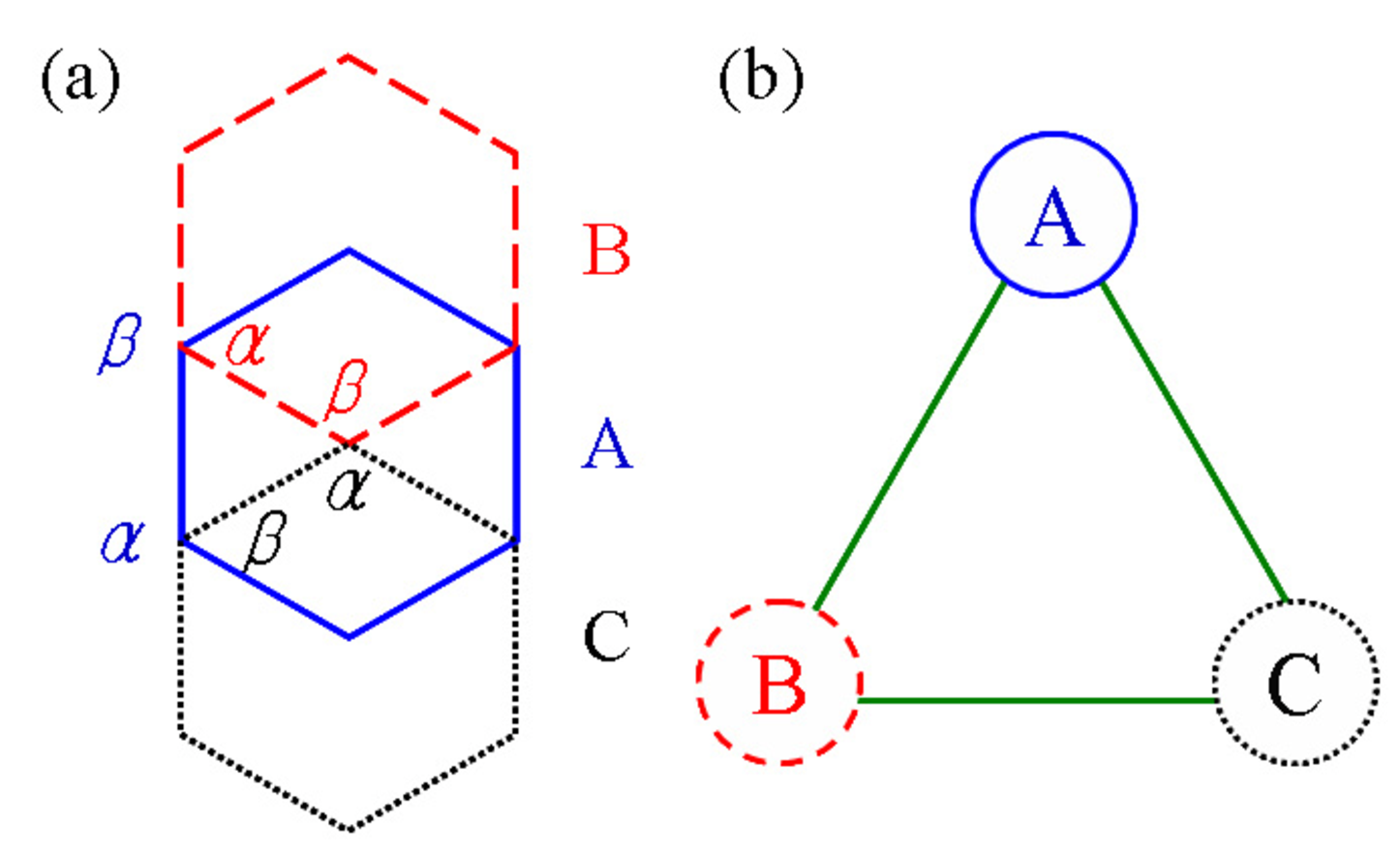}
\caption{(Color online) (a) Energetically favored stacking arrangements for graphene sheets.
The honeycomb lattice of a single sheet has two triangular sublattices, labeled by
$\alpha$ and $\beta$.  Given a starting graphene sheet, the honeycomb lattice for the next layer is 
usually positioned by displacing either $\alpha$ or $\beta$ sublattice carbon atoms along a honeycomb edge.
There are therefore in three distinct two-dimensional (2D) sheets, labeled by A, B, and C.
Representative $\alpha$ and $\beta$ sublattice positions in A, B, and C layers are identified in this illustration.
It is also possible to transform between layer types by rotating by $\pm 60^\circ$ about a 
carbon atom on one of the two sublattices.  (b) Each added layer cycles around this stacking triangle in either the
right-handed or the left-handed sense.  Reversals of the sense of this rotation tend
to increase the number of low-energy pseudospin doublets $N_D$.  In graphite, Bernal (AB) stacking
corresponds to a reversal at every step and orthorhombic (ABC) stacking corresponds to no reversals.
}
\label{fig:stack}
\end{center}
\end{figure}

\begin{figure}[htb]
\begin{center}
\includegraphics[width=0.9\linewidth]{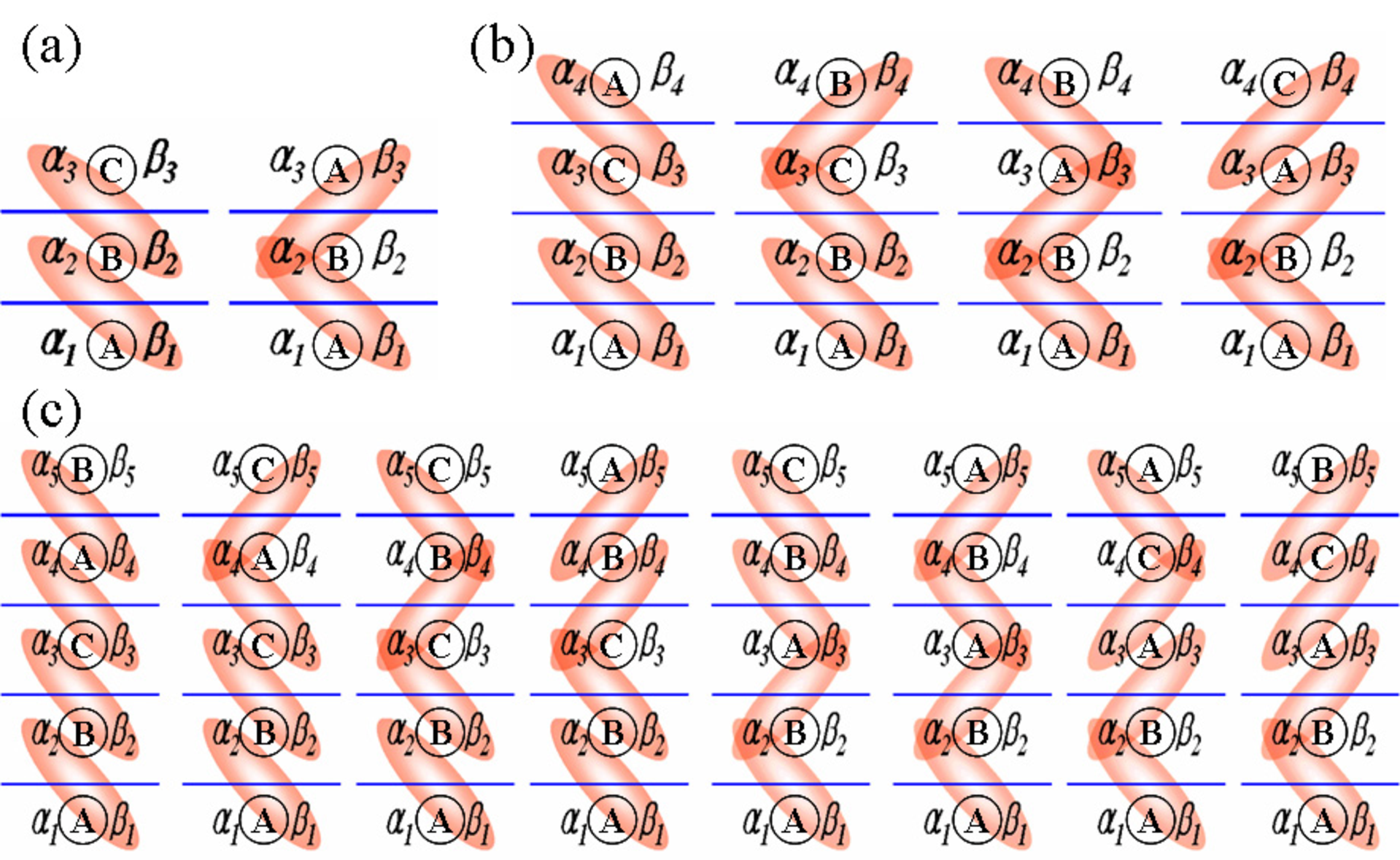}
\caption{(Color online) Stacking sequences and linkage diagrams for $N=3,4,5$ layer stacks.
The low-energy band and Landau level structures of a graphene stacks with nearest-neighbor hopping are readily read off these diagrams
as explained in the text.  Shaded ovals link
$\alpha$ and $\beta$ nearest interlayer neighbors.}
\label{fig:min_diagrams}
\end{center}
\end{figure}

When one graphene layer is placed on another, it is energetically favorable\cite{charlier1992}
for the atoms of either $\alpha$ or $\beta$ sublattices to be displaced along the honeycomb edges, as illustrated in Fig.~\ref{fig:stack}.  This stacking rule implies the three distinct but equivalent projections
(labeled A, B, and C) of the three-dimensional structure's honeycomb-lattice layers onto the $\hat{x}$-$\hat{y}$ plane
and $2^{N-2}$ distinct $N$-layer stack sequences.  When a B layer is placed on an A layer, a C layer on a B layer, or an A layer on a C layer,
the $\alpha$ sites of the upper layer are above the $\beta$ sites of the lower layer and therefore 
linked by the nearest interlayer neighbor $\pi$-orbital hopping amplitude $t_{\perp}$.
For the corresponding anticyclic stacking choices (A on B, B on C, or C on A), it is the $\beta$ sites of the upper layer and the
$\alpha$ sites of the lower layer that are linked.
All distinct $N=3$, $N=4$, and $N=5$ layer stacks are illustrated in Fig.~\ref{fig:min_diagrams}, in which we have arbitrarily labeled the first two layers starting from the bottom as A and B.  

\subsection{Energy band structure}
\subsubsection{Preliminaries}

Before analyzing energy spectrum of multilayer graphene,
let us consider the Hamiltonian of a one-band tight-binding 
model for a chain of length $N$ with near-neighbor hopping parameter
$t_{\perp}$:
\begin{equation}
H=\left(
\begin{array}{ccccccc}
0      &t_{\perp}      &0      &0      &       \\
t_{\perp}      &0      &t_{\perp}      &0      &       \\
0      &t_{\perp}      &0      &t_{\perp}      &\cdots \\
0      &0      &t_{\perp}      &0      &       \\
       &       &\cdots &       &       \\
\end{array}
\right).
\end{equation}
This Hamiltonian is important for analyzing the role of interlayer 
hopping as we explain below.

Let $\bm{a}=(a_1,...,a_N)$ be an eigenvector with an eigenvalue $\varepsilon$.
Then the eigenvalue problem reduces to the following difference equation
\begin{equation}
\varepsilon a_n = t_{\perp}( a_{n-1}+a_{n+1} ),
\end{equation}
with the boundary condition $a_0=a_{N+1}=0$.
Assuming $a_n\sim e^{i n\theta}$, it can be shown that\cite{ritger1968}
\begin{eqnarray}
\label{eq:chain}
\varepsilon_r&=&2 \, t_{\perp} \, \cos\theta_r,\nonumber  \\
{\bm a}_r &=& \sqrt{2\over N+1}(\sin\theta_r,\sin 2\theta_r,\cdots,\sin N\theta_r), 
\end{eqnarray}
where $r = 1,2,\ldots, N$ is the chain eigenvalue index and
$\theta_r=r\pi/(N+1)$.  Note that odd $N$ chains have a zero-energy eigenstate with an
eigenvector that has nonzero amplitudes, constant in magnitude and alternating in sign,
on the sublattice of the chain ends.

\subsubsection{AA stacking}
Although AA stacking is not energetically favorable, it is still interesting to consider this arrangement 
for pedagogical purposes.
In the case of AA stacking, the Hamiltonian at $K$ is given by
\begin{equation}
\label{eq:hamiltonian_AA}
H_{\rm AA}({\bm p})=\left(
\begin{array}{ccccccc}
0              &v\pi^{\dagger} &t_{\perp}      &0              &0              &0              &               \\
v\pi           &0              &0              &t_{\perp}      &0              &0              &               \\
t_{\perp}      &0              &0              &v\pi^{\dagger} &t_{\perp}      &0              &               \\
0              &t_{\perp}      &v\pi           &0              &0              &t_{\perp}      &\cdots         \\
0              &0              &t_{\perp}      &0              &0              &v\pi^{\dagger} &               \\
0              &0              &0              &t_{\perp}      &v\pi           &0              &               \\
               &               &               &\cdots         &               &               &               \\
\end{array}
\right),
\end{equation}
where $\pi=p_x+i p_y$.

As we now explain, the electronic structure of AA stacked $N$-layer graphene can be thought of 
as consisting of separate 1D chains for each wavevector in the 2D triangular lattice
Brillouin zone of a single graphene layer.
For an eigenvector $(a_1,b_1,\cdots,a_N,b_N)$ with an eigenvalue $\varepsilon$ and 
fixed 2D momentum, the difference equations in this case are
\begin{eqnarray}
\varepsilon a_n &=& t_{\perp}( a_{n-1}+a_{n+1} ) + v\pi^{\dagger} b_n,
\nonumber \\
\varepsilon b_n &=& t_{\perp}( b_{n-1}+b_{n+1} ) + v\pi a_n, 
\end{eqnarray}
with the boundary condition $a_0=a_{N+1}=b_0=b_{N+1}=0$.

Let $c_n=a_n+b_n e^{-i\phi}$ and $d_n=a_n-b_n e^{-i\phi}$ where $\phi=\tan^{-1}(p_y/p_x)$, then
\begin{eqnarray}
(\varepsilon-v|{\bm p}|) c_n &=& t_{\perp}( c_{n-1}+c_{n+1} ),
\nonumber \\
(\varepsilon+v|{\bm p}|) d_n &=& t_{\perp}( d_{n-1}+d_{n+1} ), 
\end{eqnarray}
with the same boundary condition $c_0=c_{N+1}=d_0=d_{N+1}=0$.
Thus the energy spectrum is given by
\begin{equation}
\label{eq:energy_AA}
\varepsilon^{\pm}_{r,\bm{p}}=\pm v|\bm{p}|+2 t_{\perp} \cos\left(r\pi\over{N+1}\right),
\end{equation}
where $r=1,2,\cdots,N$.
Note that for odd $N$, the $r=(N+1)/2$ mode provides two zero-energy states at ${\bm p}=0$.

\begin{figure}[h]
\includegraphics[width=0.48\linewidth]{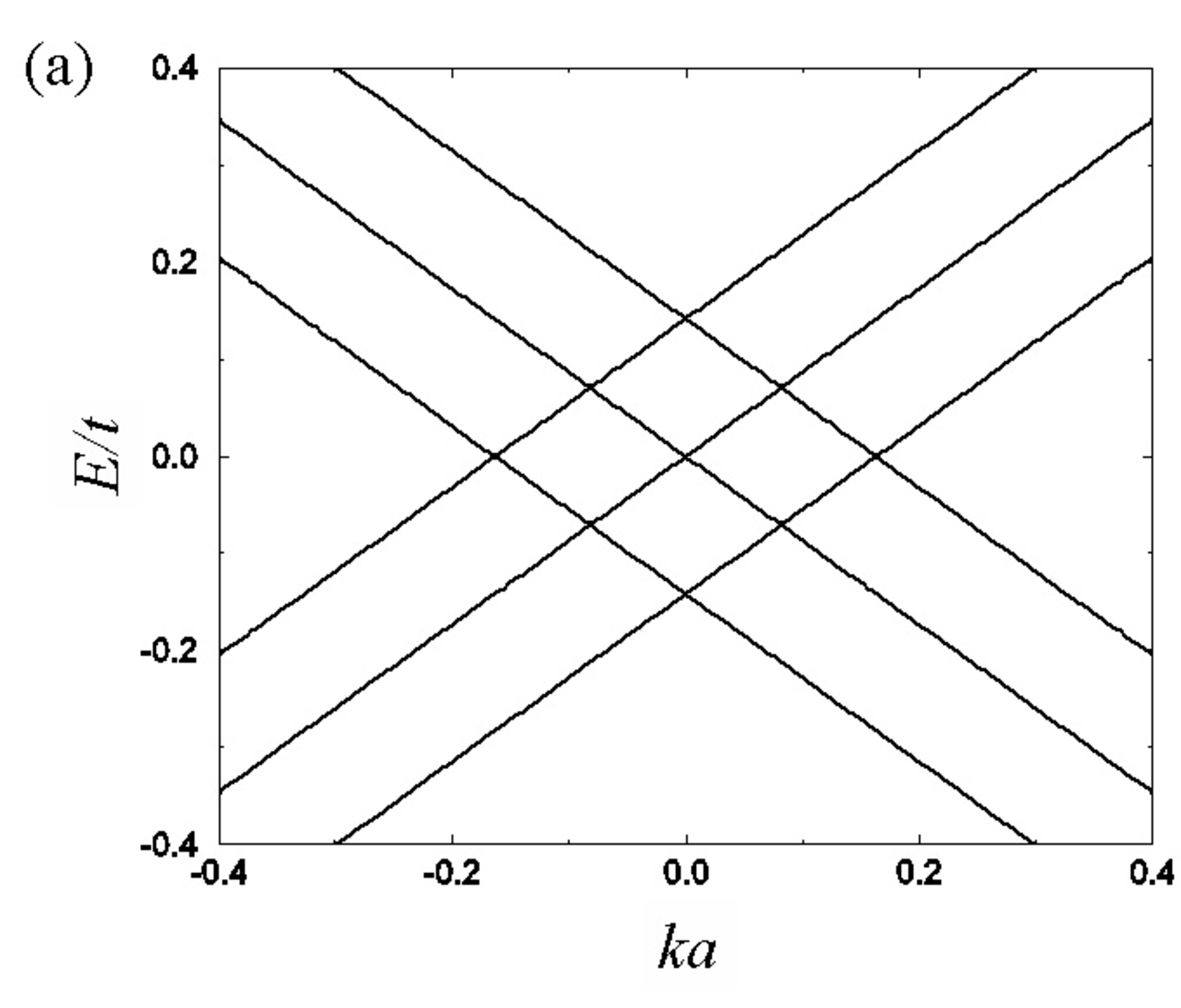}
\includegraphics[width=0.48\linewidth]{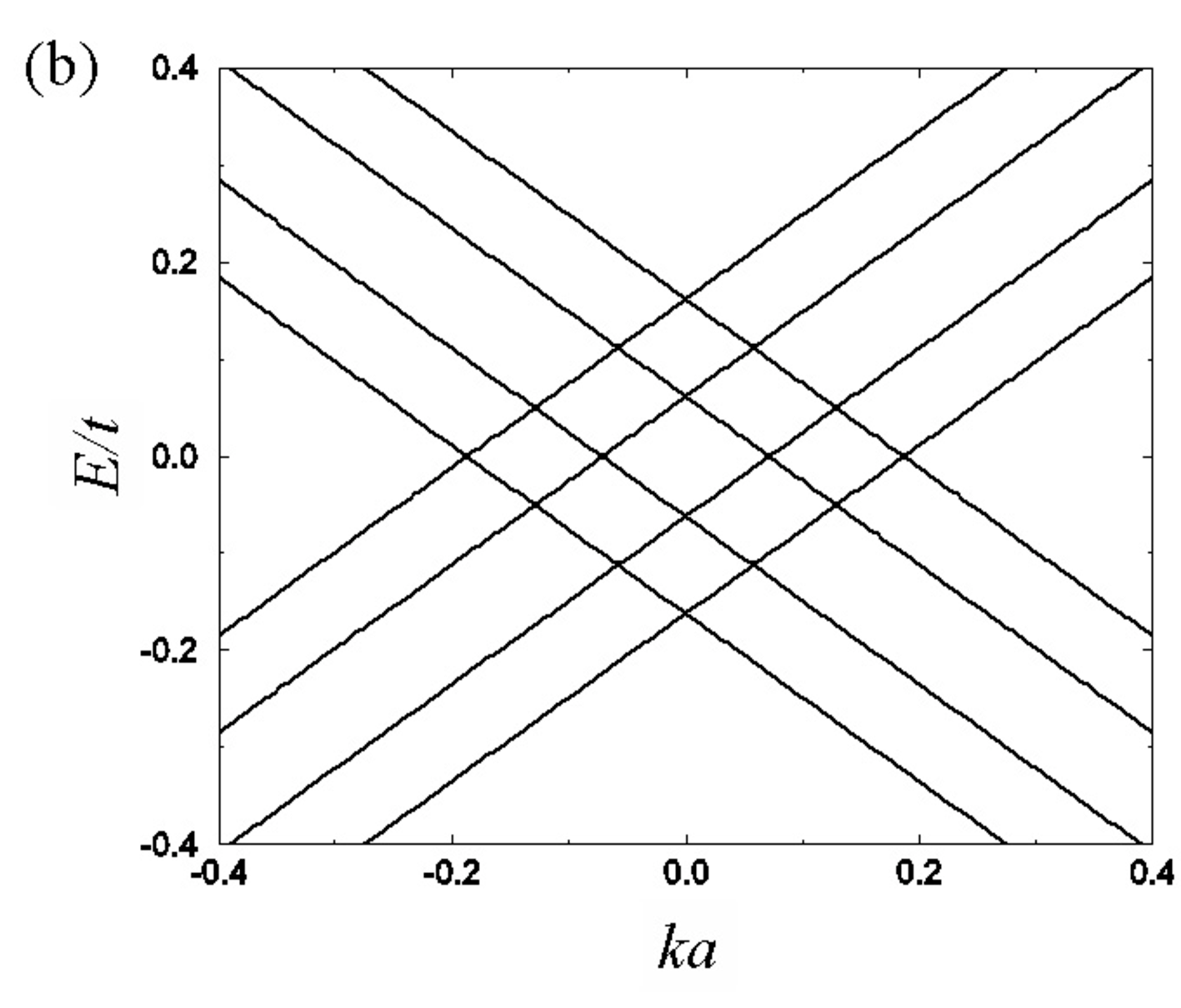}
\caption{Band structure near the $K$ point for (a) trilayer and (b) tetralayer graphene with AA stacking for nearest intralayer neighbor hopping 
$t=3$ eV and nearest interlayer neighbor hopping $t_{\perp}=0.1t$.}
\label{fig:band_AA}
\end{figure}
Figure \ref{fig:band_AA} shows the band structure of AA stacked trilayer and
tetralayer graphene near the $K$ point. Because of the hybridization between $\alpha$-$\alpha$ and $\beta$-$\beta$ sublattices in each layer, 
zero-energy states occur at momenta that are remote from the $K$ and $K'$ points.
In the following we turn our attention to stacks in which adjacent graphene layers have a relative 
rotation of 60 degrees.  As we show, in this case the zero-energy states always occur precisely at the 
Brillouin-zone corners. 

\subsubsection{AB stacking}
In the case of AB stacking, the Hamiltonian at $K$ has the following form,
\begin{equation}
\label{eq:hamiltonian_AB}
H_{\rm AB}({\bm p})=\left(
\begin{array}{ccccccc}
0              &v\pi^{\dagger} &0              &0              &0              &0              &               \\
v\pi           &0              &t_{\perp}      &0              &0              &0              &               \\
0              &t_{\perp}      &0              &v\pi^{\dagger} &0              &t_{\perp}      &               \\
0              &0              &v\pi           &0              &0              &0              &\cdots         \\
0              &0              &0              &0              &0              &v\pi^{\dagger} &               \\
0              &0              &t_{\perp}      &0              &v\pi           &0              &               \\
               &               &               &\cdots         &               &               &               \\
\end{array}
\right).
\end{equation}
We will see that the subtle difference in the Hamiltonian compared to the AA case 
changes the electronic structure in a qualitative way. 
To obtain the energy spectrum of AB stacked $N$-layer graphene,
let us consider corresponding difference equations\cite{guinea2006}:
\begin{eqnarray}
\varepsilon a_{2n-1} &=& (v\pi^{\dagger}) b_{2n-1}, \nonumber \\
\varepsilon b_{2n-1} &=& t_{\perp}( a_{2n-2}+a_{2n} )   + (v\pi) a_{2n-1}, \nonumber \\
\varepsilon a_{2n}   &=& t_{\perp}( b_{2n-1}+b_{2n+1} ) + (v\pi^{\dagger}) b_{2n}, \nonumber \\
\varepsilon b_{2n}   &=& (v\pi) a_{2n}, 
\end{eqnarray}
with the boundary condition $a_0=a_{N+1}=b_0=b_{N+1}=0$.

Let $c_{2n-1}=b_{2n-1}$ and $c_{2n}=a_{2n}$, then the difference equations reduce to
\begin{equation}
(\varepsilon-v^2|{\bm p}|^2/\varepsilon) c_n = t_{\perp}( c_{n-1}+c_{n+1} ),
\end{equation}
with the boundary condition $c_0=c_{N+1}=0$.
Then the energy spectrum is given by
\begin{equation}
\varepsilon-v^2|{\bm p}|^2/\varepsilon=2 t_{\perp}\cos\left(r\pi\over{N+1}\right),
\end{equation}
where $r=1,2,\cdots,N$. Thus
\begin{equation}
\label{eq:energy_AB}
\varepsilon^{\pm}_{r,\bm{p}}=t_{\perp}\cos\left(r\pi\over{N+1}\right)\pm \sqrt{v^2|\bm{p}|^2+t_{\perp}^2\cos^2\left(r\pi\over{N+1}\right)}.
\end{equation}

Note that relativistic energy spectrum for a particle with the momentum $\bm{p}$ and mass $m$ is given by
\begin{equation}
\varepsilon_{\bm p}=\sqrt{|{\bm p}|^2 c^2+m^2 c^4}.
\end{equation}
Thus we can identify $m_r v^2=\left|t_{\perp}\cos\left(r\pi\over{N+1}\right)\right|$
as the effective mass for mode $r$.

For a massive mode with mass $m_r$, the low-energy spectrum is given by
\begin{equation}
\label{eq:energy_AB_eff_massive}
\varepsilon_{r,\bm p} \approx
\begin{cases}
+{{\bm p}^2 \over 2m_r} & \text{if $t_{\perp} \cos\left(r\pi\over{N+1}\right)<0$}, \\
-{{\bm p}^2 \over 2m_r} & \text{if $t_{\perp} \cos\left(r\pi\over{N+1}\right)>0$}.
\end{cases}
\end{equation}
For odd $N$, the mode with $r=(N+1)/2$ is massless and 
its energy is given by
\begin{equation}
\label{eq:energy_AB_eff_massless}
\varepsilon_{\bm p}^{\pm}\approx\pm v|{\bm p}|.
\end{equation}
For even $N$, all $N$ modes are massive at low energies.
Therefore, the low-energy spectrum with odd number of layers
is a combination of one massless Dirac mode and $N-1$ massive Dirac modes,
whereas the low-energy spectrum with even number of layers
is composed of only massive Dirac modes.

\begin{figure}[t]
\includegraphics[width=0.48\linewidth]{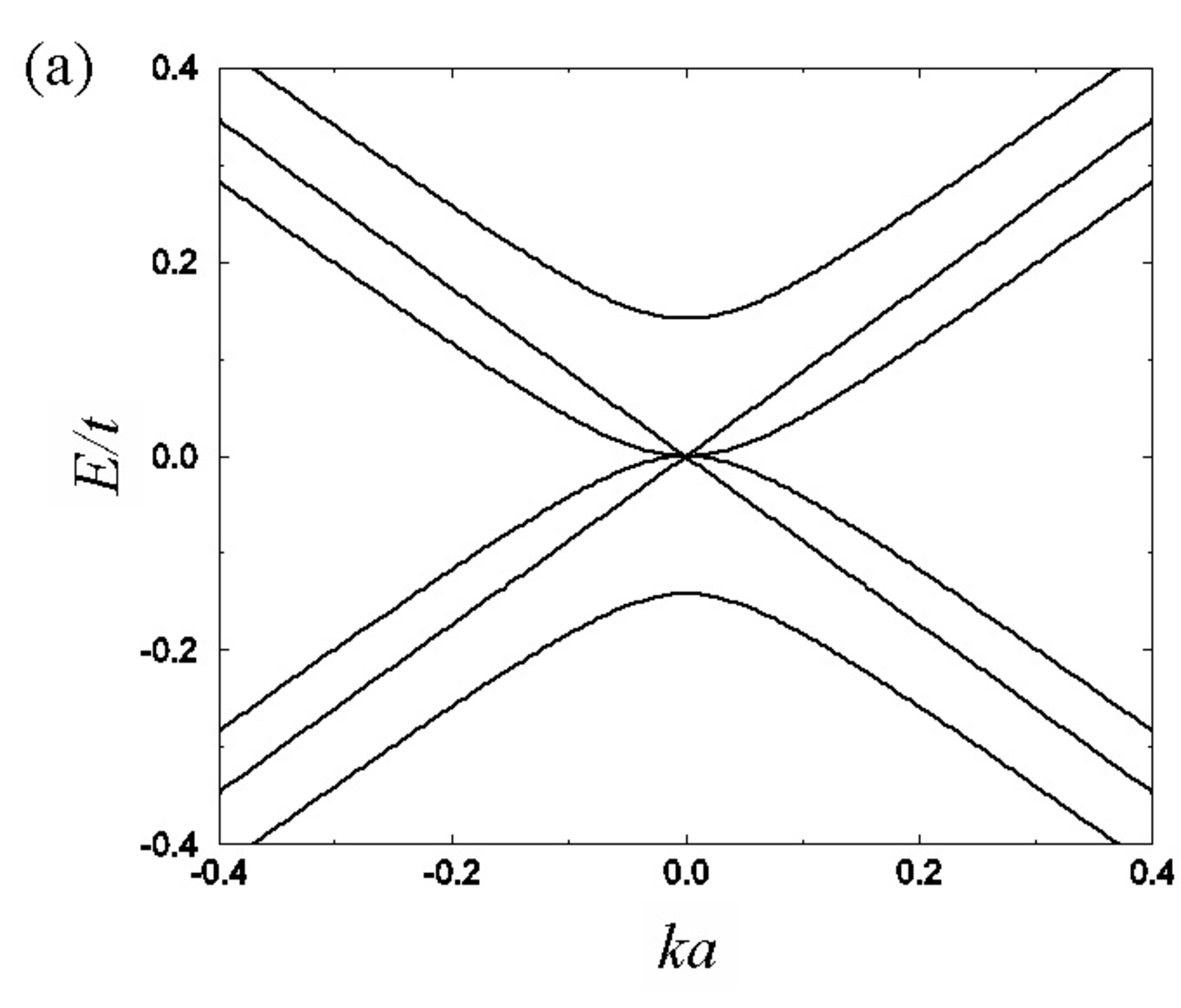}
\includegraphics[width=0.48\linewidth]{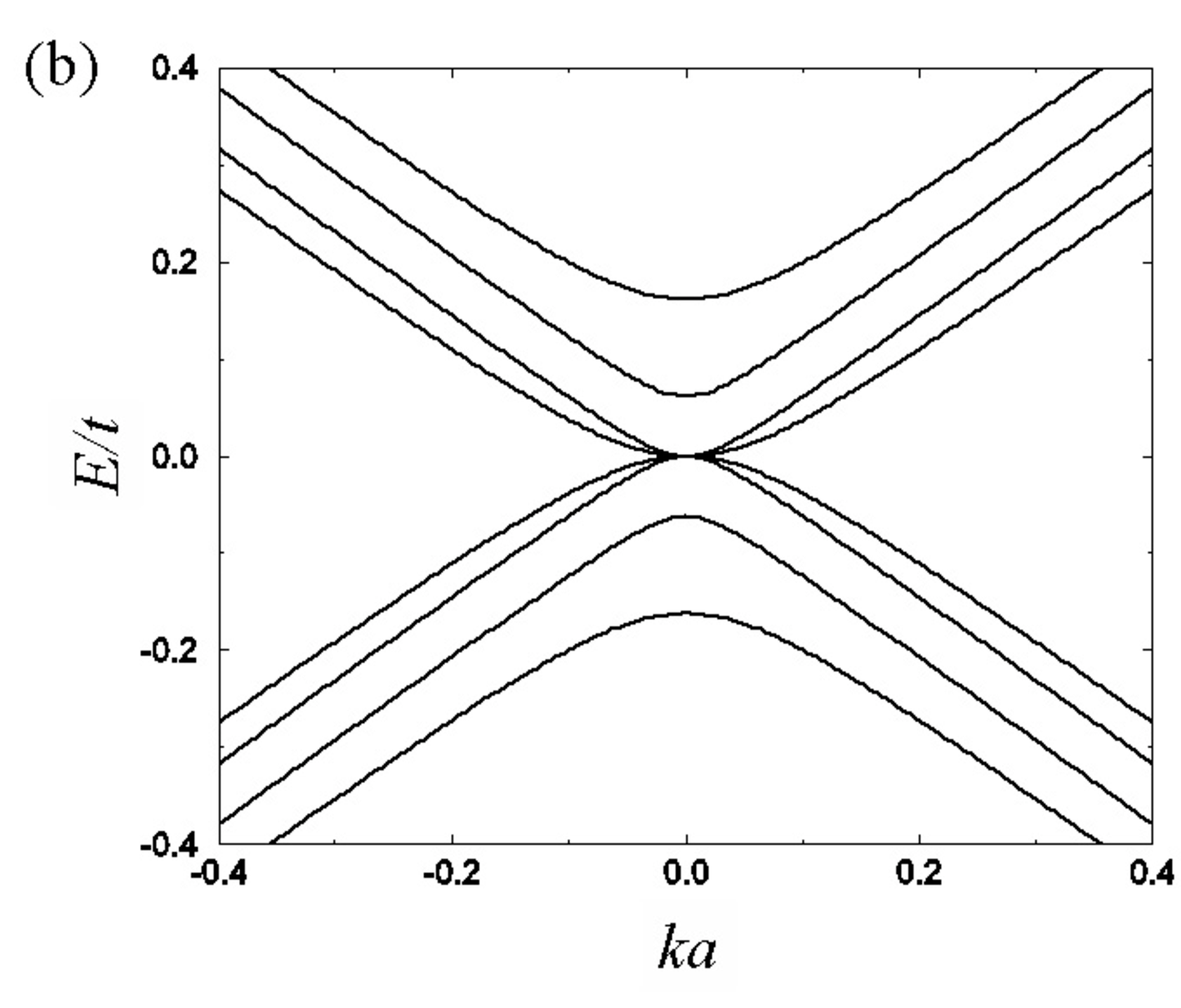}
\caption{Band structure near the $K$ point for (a) trilayer and (b) tetralayer graphene with AB stacking
for nearest intralayer neighbor hopping $t=3$ eV and nearest interlayer neighbor hopping $t_{\perp}=0.1t$.}
\label{fig:band_AB}
\end{figure}
Figure \ref{fig:band_AB} shows the band structure of AB stacked trilayer and
tetralayer graphene near the $K$ point. 
As discussed earlier, the trilayer has one massless mode and two massive modes, while the tetralayer has all massive modes at low energies.
Note that at ${\bm p}=0$, each massless mode gives two zero energies while
each massive mode gives one zero energy.
Therefore, for odd $N$, there are $2+ (N-1)=N+1$ zero-energy states 
while for even $N$, there are $N$ zero-energy states.

\subsubsection{ABC stacking}
In the case of ABC stacking, the Hamiltonian at $K$ is given by
\begin{equation}
\label{eq:hamiltonian_ABC}
H_{\rm ABC}({\bm p})=\left(
\begin{array}{ccccccc}
0              &v\pi^{\dagger} &0              &0              &0              &0              &               \\
v\pi           &0              &t_{\perp}      &0              &0              &0              &               \\
0              &t_{\perp}      &0              &v\pi^{\dagger} &0              &0              &               \\
0              &0              &v\pi           &0              &t_{\perp}      &0              &\cdots         \\
0              &0              &0              &t_{\perp}      &0              &v\pi^{\dagger} &               \\
0              &0              &0              &0              &v\pi           &0              &               \\
               &               &               &\cdots         &               &               &               \\
\end{array}
\right).
\end{equation}
Unfortunately for ABC stacking, there do not exist low-order difference equations
with a simple boundary condition.
Instead we can easily derive a low-energy effective Hamiltonian.
Surprisingly, it turns out that ABC stacked $N$-layer graphene is described by
$N$-chiral 2D electron system. (More detailed discussion for the effective theory of arbitrarily stacked graphene will be presented in \S\ref{sec:chiral_decomposition}.)  

It is important to recognize that in ABC stacking, there is vertical hopping
between all the lower layer $\beta$ sites and all the upper layer $\alpha$ sites.
For $\pi=0$ each $\alpha$-$\beta$ pair forms a symmetric-antisymmetric doublet with 
energies $\pm t_{\perp}$,
leaving the bottom $\alpha_1$ and top $\beta_N$ sites as the only low-energy states.
This behavior is readily understood from the stacking diagrams,
in Fig.~\ref{fig:min_diagrams}.
It is possible to construct a $2\times 2$ $\pi$-dependent 
low-energy effective Hamiltonian for the low-energy part of the spectrum using 
perturbation theory.  The same procedure can then be extended to arbitrary stacking 
sequences. 

The simplest example is bilayer graphene\cite{mccann2006}.  Low and high energy
subspaces are identified by finding the spectrum at $\pi=0$ and identifying all the 
zero-energy eigenstates.  The intralayer tunneling term, which is proportional to $\pi$, 
couples low and high energy states.  Using degenerate state perturbation theory, the 
effective Hamiltonian in the low energy space is given to leading (2nd) order in $\pi$ by 
\begin{equation}
H_2^{eff}({\bm p})=-\left(
\begin{array}{cc}
0                &  {(\pi^{\dagger})^2 \over 2m}  \\
{(\pi)^2 \over 2m} &  0                            \\
\end{array}
\right)=
-t_{\perp}\left(
\begin{array}{cc}
0          &  (\nu^{\dagger})^2   \\
(\nu)^2  &  0                      \\
\end{array}
\right),
\end{equation}
where we have used a $(\alpha_1,\beta_2)$ basis, $m=t_{\perp}/2v^2$ and $\nu={v\pi / t_{\perp}}$.
In the same way we find that the effective Hamiltonian of ABC stacked $N$-layer graphene is given by
\begin{equation}
\label{eq:hamiltonian_ABC_eff}
H_N^{eff}({\bm p})=-t_{\perp}\left(
\begin{array}{cc}
0          & (\nu^{\dagger})^N \\
(\nu)^N & 0                    \\
\end{array}
\right),
\end{equation}
using a $(\alpha_1,\beta_N)$ basis.
The leading correction appears at order $N$ in $\pi$ because the unperturbed high-energy states 
are localized on a $(\beta_i,\alpha_{i+1})$ pair and the perturbation is intralayer tunneling.  
Note that we have for mathematical convenience chosen a gauge in which the single-layer Hamiltonian is 
\begin{equation}
H_1^{eff}({\bm p})=-\left(
\begin{array}{cc}
0          &  v \pi^{\dagger} \\
v \pi      &  0               \\
\end{array}
\right).
\end{equation}

We can prove Eq.~(\ref{eq:hamiltonian_ABC_eff}) by the mathematical induction method.
Imagine that we add one more layer on top of $N$-layer graphene with ABC stacking.
Then the combined Hamiltonian is given by
\begin{equation}
H_{N+1}^{eff}({\bm p})=-t_{\perp}\left(
\begin{array}{cccc}
0          & (\nu^{\dagger})^N & 0      & 0                \\
(\nu)^N & 0                    & -1     & 0                \\
0          & -1                   & 0      & \nu^{\dagger} \\
0          & 0                    & \nu & 0                \\
\end{array}
\right),
\end{equation}
using a $(\alpha_1,\beta_N,\alpha_{N+1},\beta_{N+1})$ basis.

Let $P$ be a low-energy subspace spanned by $(\alpha_1,\beta_{N+1})$
and $Q$ be a high-energy subspace spanned by $(\alpha_{N+1},\beta_N)$.
Note that the effective Hamiltonian can be derived using the degenerate state perturbation theory\cite{sakurai1994},
\begin{equation}
\label{eq:effective}
H_{eff}\approx H_{PP}-H_{PQ}{1\over {H_{QQ}}} H_{QP}.
\end{equation}
Here the projected Hamiltonian matrices to $P$ and $Q$ subspace are given by
\begin{equation}
H_{QQ}({\bm p})=t_{\perp}\left(
\begin{array}{cc}
0  &  1 \\
1  &  0  \\
\end{array}
\right), \,
H_{PQ}({\bm p})=-t_{\perp}\left(
\begin{array}{cc}
0       &  (\nu^{\dagger})^N \\
\nu  &  0                    \\
\end{array}
\right),
\end{equation}
and $H_{PP}({\bm p})=0$.
Thus we can easily show that,
\begin{equation}
H_{N+1}^{eff}({\bm p})\approx-t_{\perp}\left(
\begin{array}{cc}
0          & (\nu^{\dagger})^{N+1} \\
(\nu)^{N+1} & 0                    \\
\end{array}
\right),
\end{equation}
which proves Eq.~(\ref{eq:hamiltonian_ABC_eff}).
The corresponding energy spectrum in Eq.~(\ref{eq:hamiltonian_ABC_eff}) is given by
\begin{equation}
\label{eq:energy_ABC_eff}
\varepsilon_{eff,{\bm p}}^{\pm}=\pm t_{\perp}\left(v|{\bm p}|\over t_{\perp}\right)^N.
\end{equation}

\begin{figure}[h]
\includegraphics[width=0.48\linewidth]{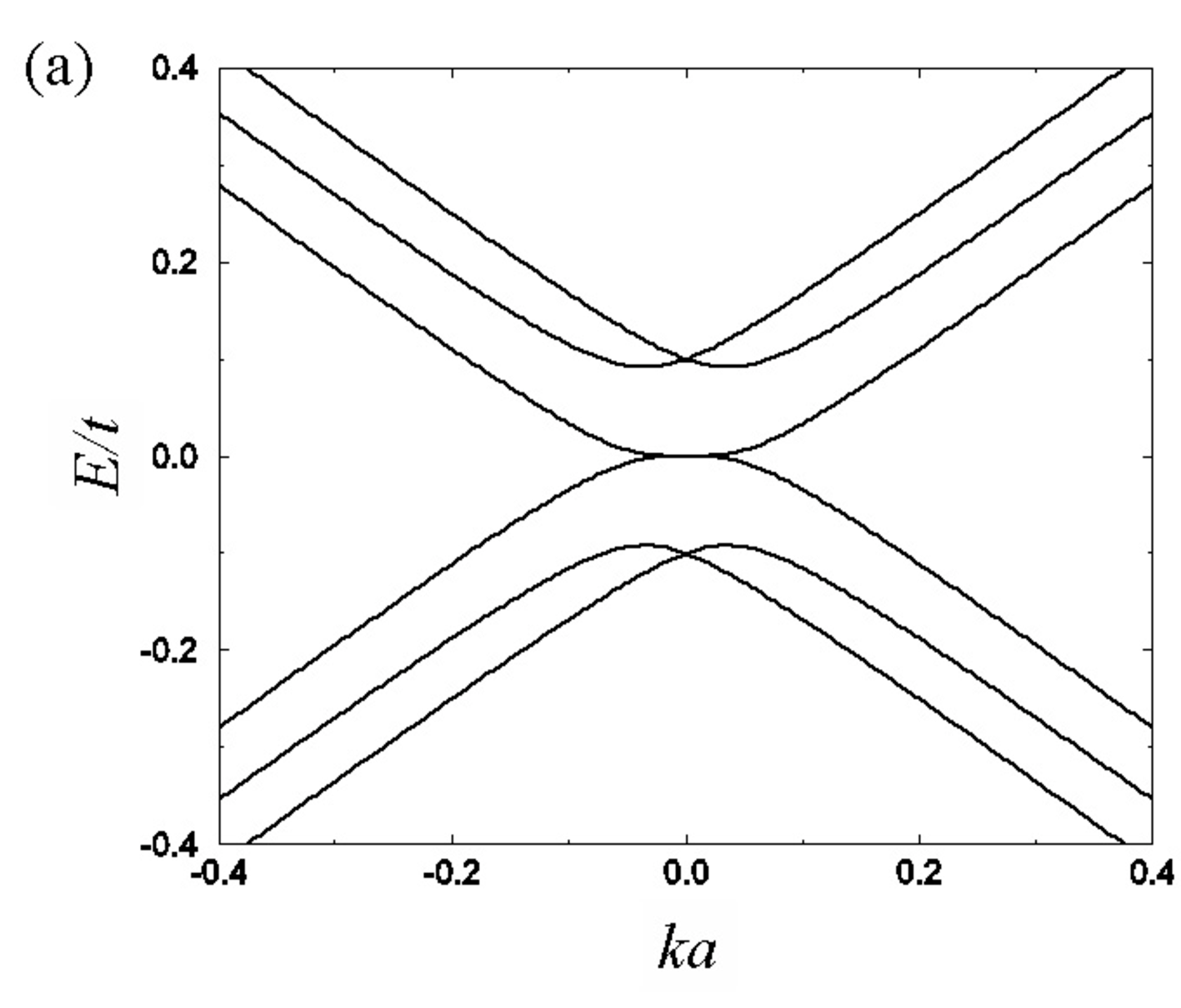}
\includegraphics[width=0.48\linewidth]{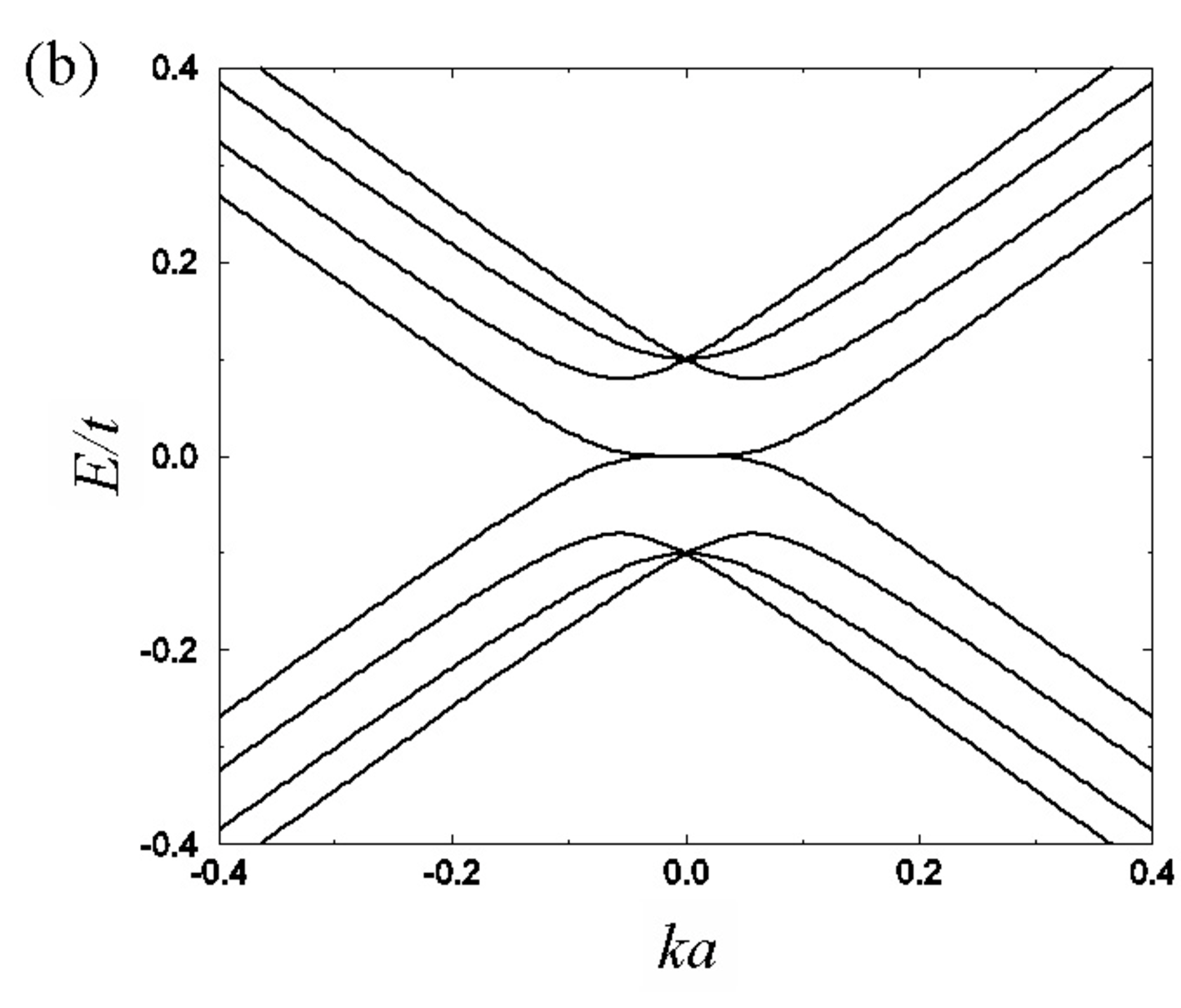}
\caption{Band structure near the $K$ point for (a) trilayer and (b) tetralayer graphene with ABC stacking
for nearest intralayer neighbor hopping $t=3$ eV and nearest interlayer neighbor hopping $t_{\perp}=0.1t$.}
\label{fig:band_ABC}
\end{figure}
Figure \ref{fig:band_ABC} shows the band structure of ABC stacked trilayer and
tetralayer graphene near the $K$ point.
Note that at ${\bm p}=0$, there are only two zero energy states no matter how thick the stack is.

\subsubsection{Arbitrary stacking}
It is easy to generalize the previous discussion to construct the Hamiltonian for an arbitrarily stacked multilayer graphene system.
\begin{figure}[h]
\includegraphics[width=0.48\linewidth]{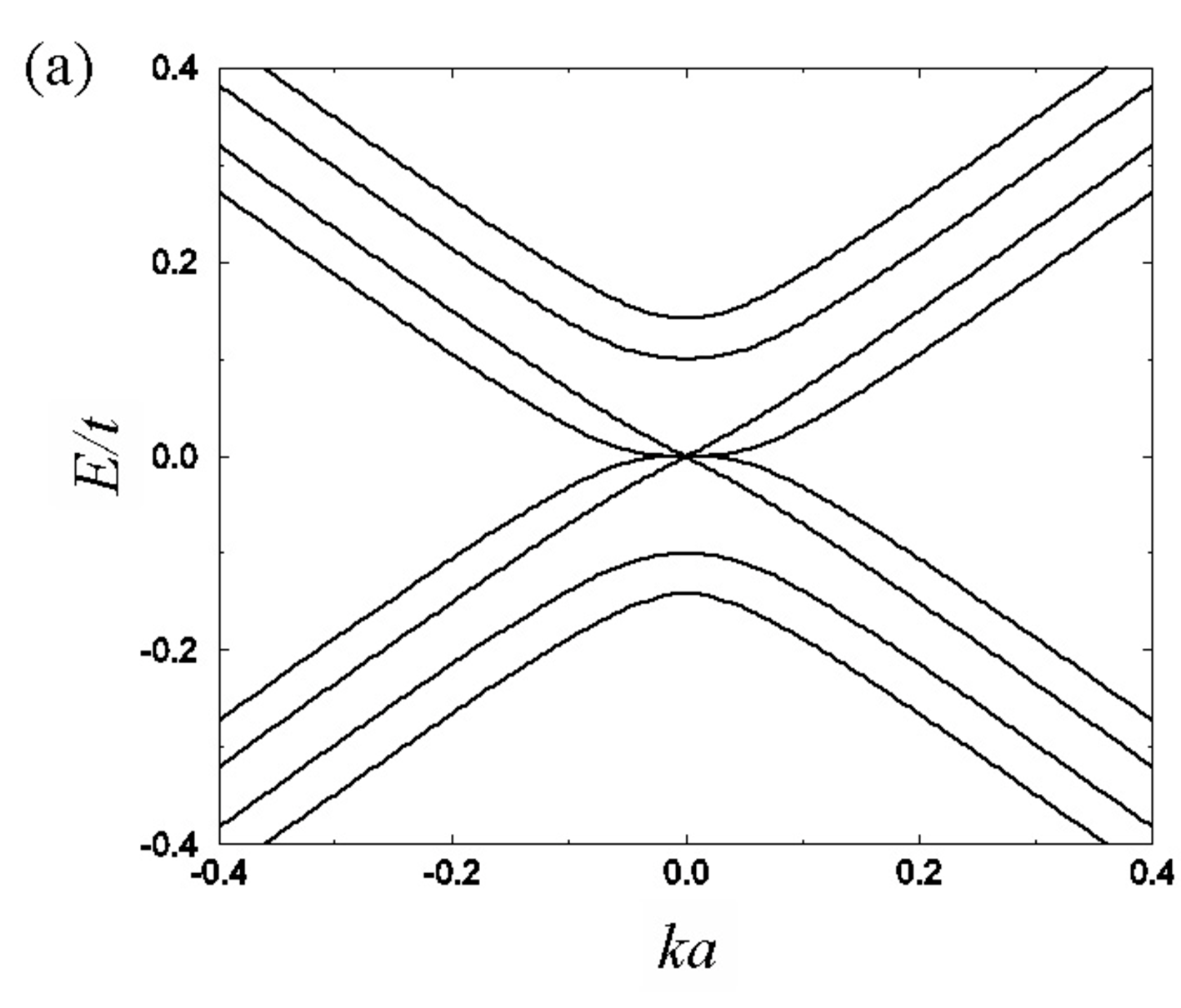}
\includegraphics[width=0.48\linewidth]{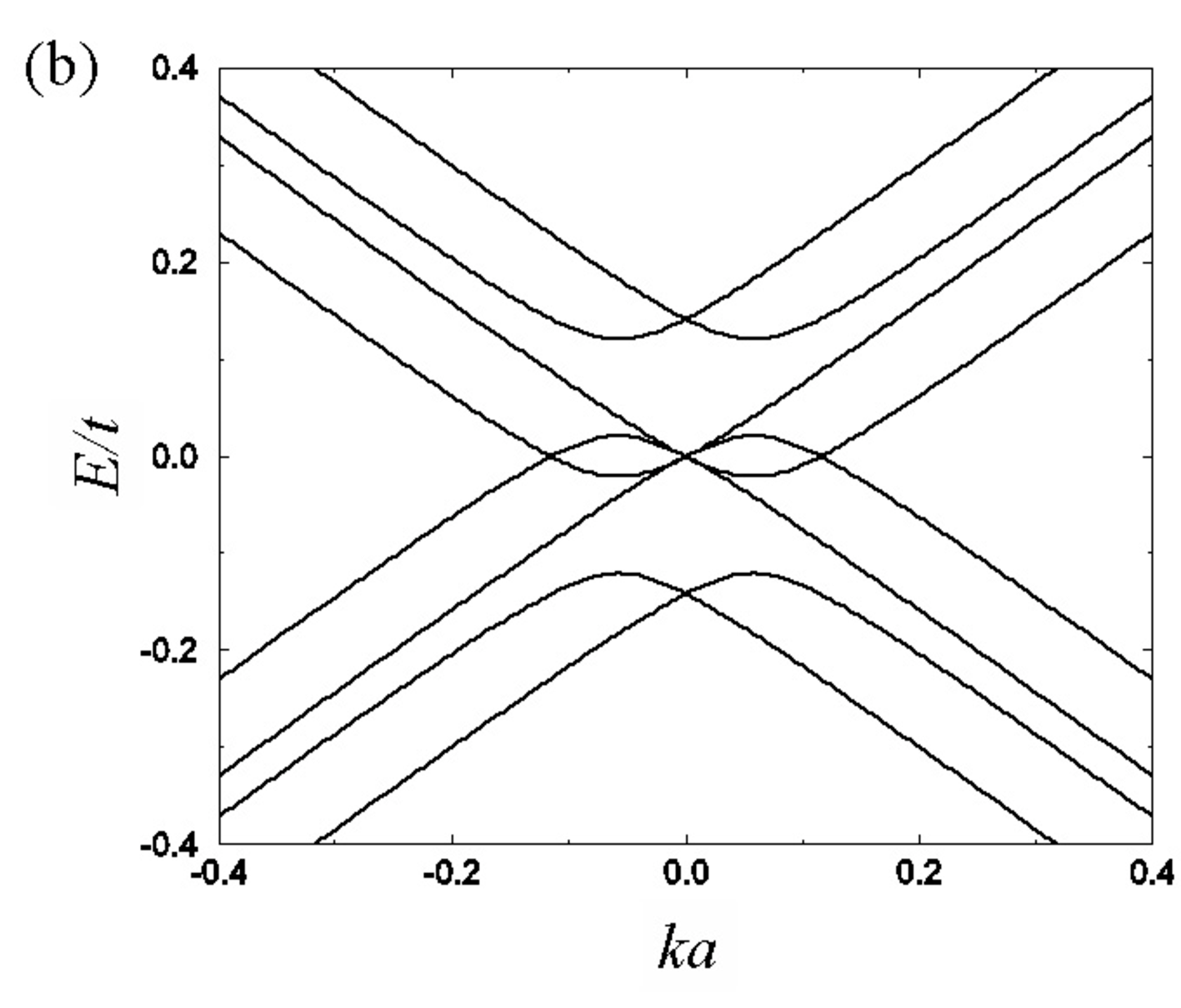}
\caption{Band structure near the $K$ point for tetralayer graphene with (a) ABCB stacking and (b) ABBC stacking 
for nearest intralayer neighbor hopping $t=3$ eV and nearest interlayer neighbor hopping $t_{\perp}=0.1t$.}
\label{fig:band_arbitrary}
\end{figure}
Figure \ref{fig:band_arbitrary} shows the band structure of ABCB stacked tetralayer graphene and ABBC stacked tetralayer graphene near the $K$ point.
For the ABCB stacked tetralayer graphene, the low-energy spectrum looks like a superposition
of a linear dispersion and a cubic one. For the ABBA stacked tetralayer graphene, 
zero energies appear not only at the Dirac point but also away from it.
A more detailed low-energy spectrum analysis will be presented in \S\ref{sec:chiral_decomposition}.

\subsection{Landau level spectrum}

\subsubsection{Preliminaries}
In the presence of a magnetic field ${\bm B}=B\hat{z}$, a Hamiltonian
is modified by ${\bm p}\rightarrow {\bm p}+{e\over c}{\bm A}$,
where ${\bm A}$ is the vector potential with ${\bm B}=\nabla\times {\bm A}$.
The quantum Hamiltonian is most easily diagonalized by introducing
raising and lowering operators, $a=l\pi^{\dagger}/{\sqrt 2}\hbar$ and $a^{\dagger}=l\pi/{\sqrt 2}\hbar$ substitution,
where $l=\sqrt{\hbar c/ e|B|}$, and noting that $[a,a^{\dagger}]=1$. 
We can then expand the wavefunction amplitude on each sublattice of each layer in terms
of parabolic band Landau level states $\left|n\right>$ which are eigenstates of the $a^{\dagger} a$. 
For many Hamiltonians, including those studied here, the Hamiltonian can be block diagonalized by fixing the parabolic band
Landau-level offset between different sublattices and between different layers.  
This procedure is familiar from theories of Landau-level structure in other multiband $\bm{k} \cdot \bm{p}$ theories. 

\subsubsection{AA stacking}
In the case of AA stacking, let us choose the $n$-th Landau level basis at $K$ as\\
$(\alpha_{1,n-1},\beta_{1,n},\cdots,\alpha_{N,n-1},\beta_{N,n})$.
Then Eq.~(\ref{eq:hamiltonian_AA}) reduces to
\begin{equation}
\label{eq:hamiltonian_AA_LL}
H_{\rm AA}(n)=\left(
\begin{array}{ccccccc}
0              &\varepsilon_n     &t_{\perp}      &0              &0              &0              &               \\
\varepsilon_n     &0              &0              &t_{\perp}      &0              &0              &               \\
t_{\perp}      &0              &0              &\varepsilon_n     &t_{\perp}      &0              &               \\
0              &t_{\perp}      &\varepsilon_n     &0              &0              &t_{\perp}      &\cdots         \\
0              &0              &t_{\perp}      &0              &0              &\varepsilon_n     &               \\
0              &0              &0              &t_{\perp}      &\varepsilon_n     &0              &               \\
               &               &               &\cdots         &               &               &               \\
\end{array}
\right),
\end{equation}
where $\varepsilon_n=\sqrt{2 n}\hbar v /l$.
Note that 2D Landau level states with a negative index do not exist so
the corresponding basis states and matrix elements are understood as being absent in the matrix block.
Thus $H_{\rm AA}(n=0)$ is a $N\times N$ matrix, while $H_{\rm AA}(n>0)$ is a $2N\times 2N$ matrix.

By diagonalizing Eq.~(\ref{eq:hamiltonian_AA_LL}) using the difference equation method,
we can obtain the exact Landau level spectrum.
For $n>0$, Landau levels are given by
\begin{equation}
\varepsilon^{\pm}_{r,n}=\pm \varepsilon_n+2 t_{\perp} \cos\left(r\pi\over{N+1}\right),
\end{equation}
where $r=1,2,\cdots,N$.
Note that for $n=0$, Landau levels are given by $\varepsilon_{r,0}=2 t_{\perp} \cos\left(r\pi\over{N+1}\right)$.
Thus for odd $N$, there exists one ($B$-independent) zero-energy Landau level at $r=(N+1)/2$.

\begin{figure}[t]
\includegraphics[width=0.48\linewidth]{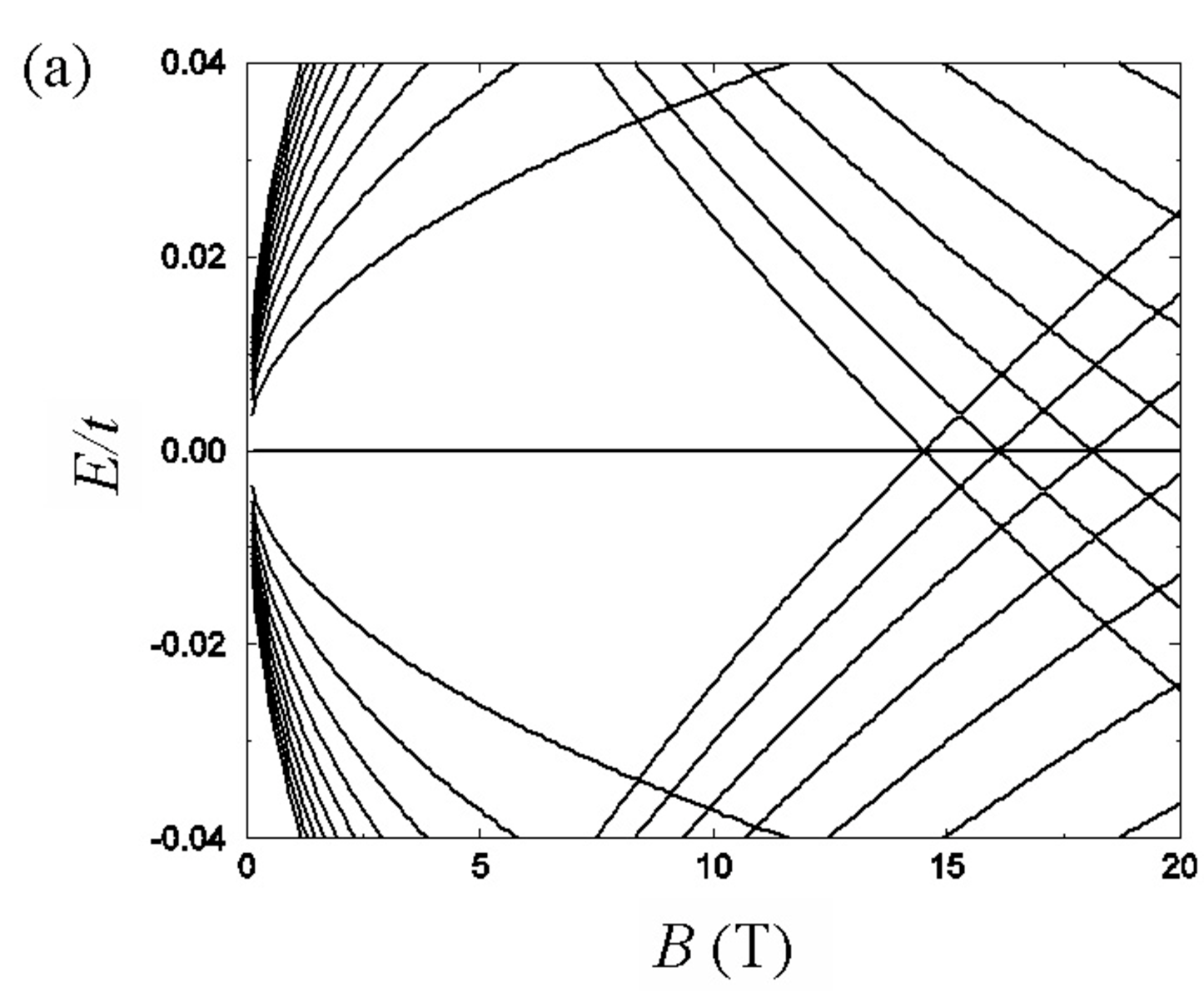}
\includegraphics[width=0.48\linewidth]{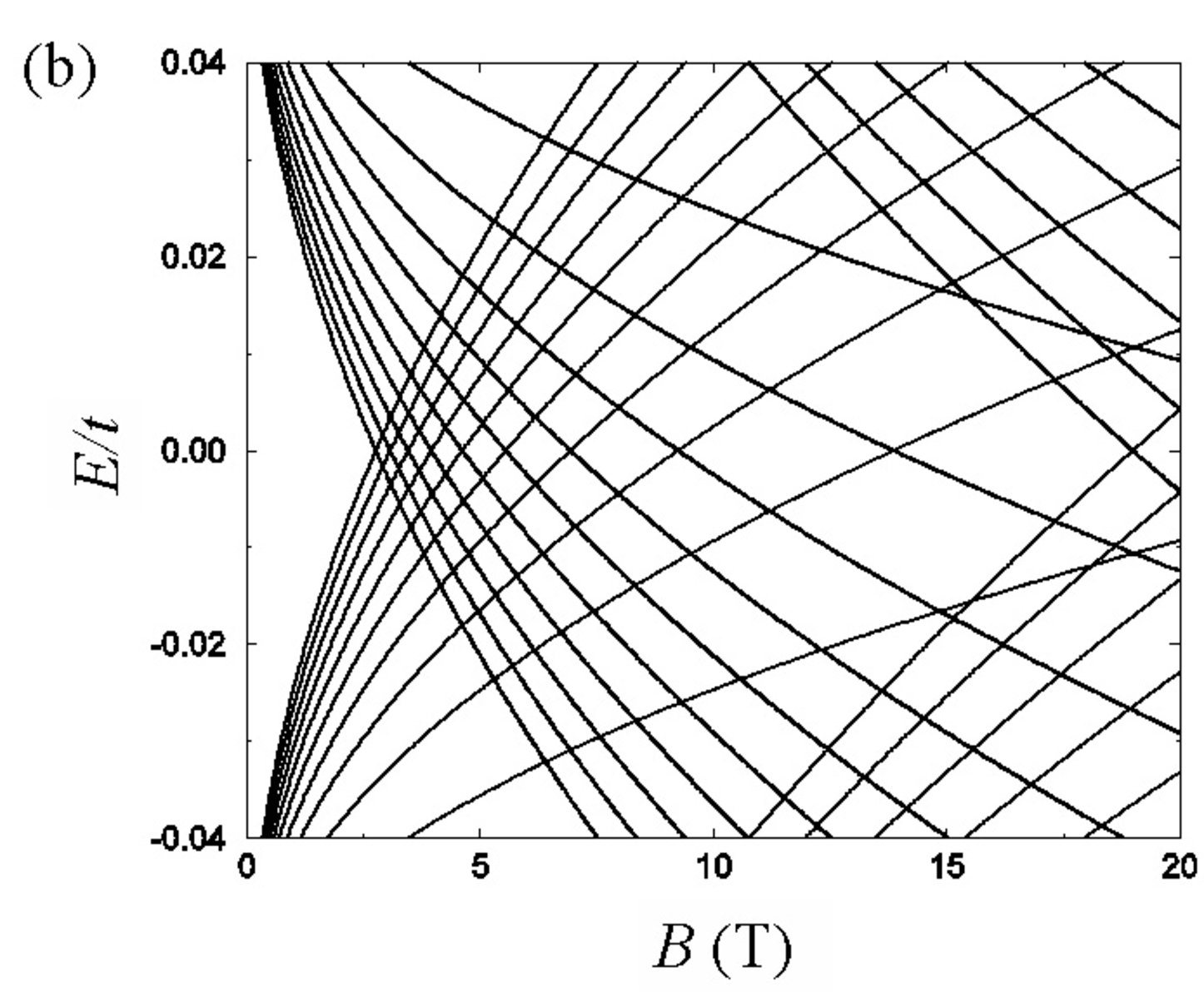}
\caption{Landau levels of (a) trilayer and (b) tetralayer graphene 
with AA stacking for nearest intralayer neighbor hopping $t=3$ eV and nearest interlayer neighbor hopping $t_{\perp}=0.1t$. Landau levels were shown up to $n=10$.}
\label{fig:LL_AA}
\end{figure}
Figure \ref{fig:LL_AA} shows the Landau levels of AA stacked trilayer and
tetralayer graphene as a function of magnetic fields.
For the trilayer, there is one zero-energy Landau level, while for the tetralayer,
there is no zero-energy Landau level.
Note that there are Landau levels crossing the zero-energy line in AA stacking.

\subsubsection{AB stacking}
In the case of AB stacking, a proper choice of the $n$-th Landau level basis at $K$ is
$(\alpha_{1,n-1},\beta_{1,n},\alpha_{2,n},\beta_{2,n+1},\alpha_{3,n-1},\beta_{3,n},\alpha_{4,n},\beta_{4,n+1},\cdots)\,$ 
such that all the interlayer hopping terms are contained in the $n$-th Landau level Hamiltonian.
Then Eq.~(\ref{eq:hamiltonian_AB}) reduces to
\begin{equation}
\label{eq:hamiltonian_AB_LL}
H_{\rm AB}(n)=\left(
\begin{array}{ccccccc}
0              &\varepsilon_n     &0              &0              &0              &0              &               \\
\varepsilon_n     &0              &t_{\perp}      &0              &0              &0              &               \\
0              &t_{\perp}      &0              &\varepsilon_{n+1} &0              &t_{\perp}      &               \\
0              &0              &\varepsilon_{n+1} &0              &0              &0              &\cdots         \\
0              &0              &0              &0              &0              &\varepsilon_n     &               \\
0              &0              &t_{\perp}      &0              &\varepsilon_n     &0              &               \\
               &               &               &\cdots         &               &               &               \\
\end{array}
\right),
\end{equation}
where $\varepsilon_n=\sqrt{2 n}\hbar v /l$. 
As discussed earlier, special care should be given for states with a negative index.

For the Hamiltonian in Eq.~(\ref{eq:hamiltonian_AB_LL}), there do not exist corresponding difference equations with a proper boundary condition, 
thus cannot be diagonalized analytically.  From Eq.~(\ref{eq:energy_AB_eff_massive}), however, 
we can find the low-energy Landau levels for massive mode with mass $m_r$ as
\begin{equation}
\label{eq:LL_massive}
\varepsilon_{r,n} \approx
\begin{cases}
+\hbar \omega_r \sqrt{n(n+1)} & \text{if $t_{\perp} \cos\left(r\pi\over{N+1}\right)<0$}, \\
-\hbar \omega_r \sqrt{n(n+1)} & \text{if $t_{\perp} \cos\left(r\pi\over{N+1}\right)>0$},
\end{cases}
\end{equation}
where $\omega_r=e|B|/m_r c$ and $r=1,2,\cdots,N$, which is proportional to $B$.
These equations apply at small $B$, just as the $B=0$ limiting low-energy 
dispersions applied at small momentum $\pi$.
For the massless mode, from Eq.~(\ref{eq:energy_AB_eff_massless}) Landau levels are given by
\begin{equation}
\label{eq:LL_massless}
\varepsilon^{\pm}_{n}=\pm\varepsilon_n,
\end{equation}
which is proportional to $B^{1/2}$.

\begin{figure}[h]
\includegraphics[width=0.48\linewidth]{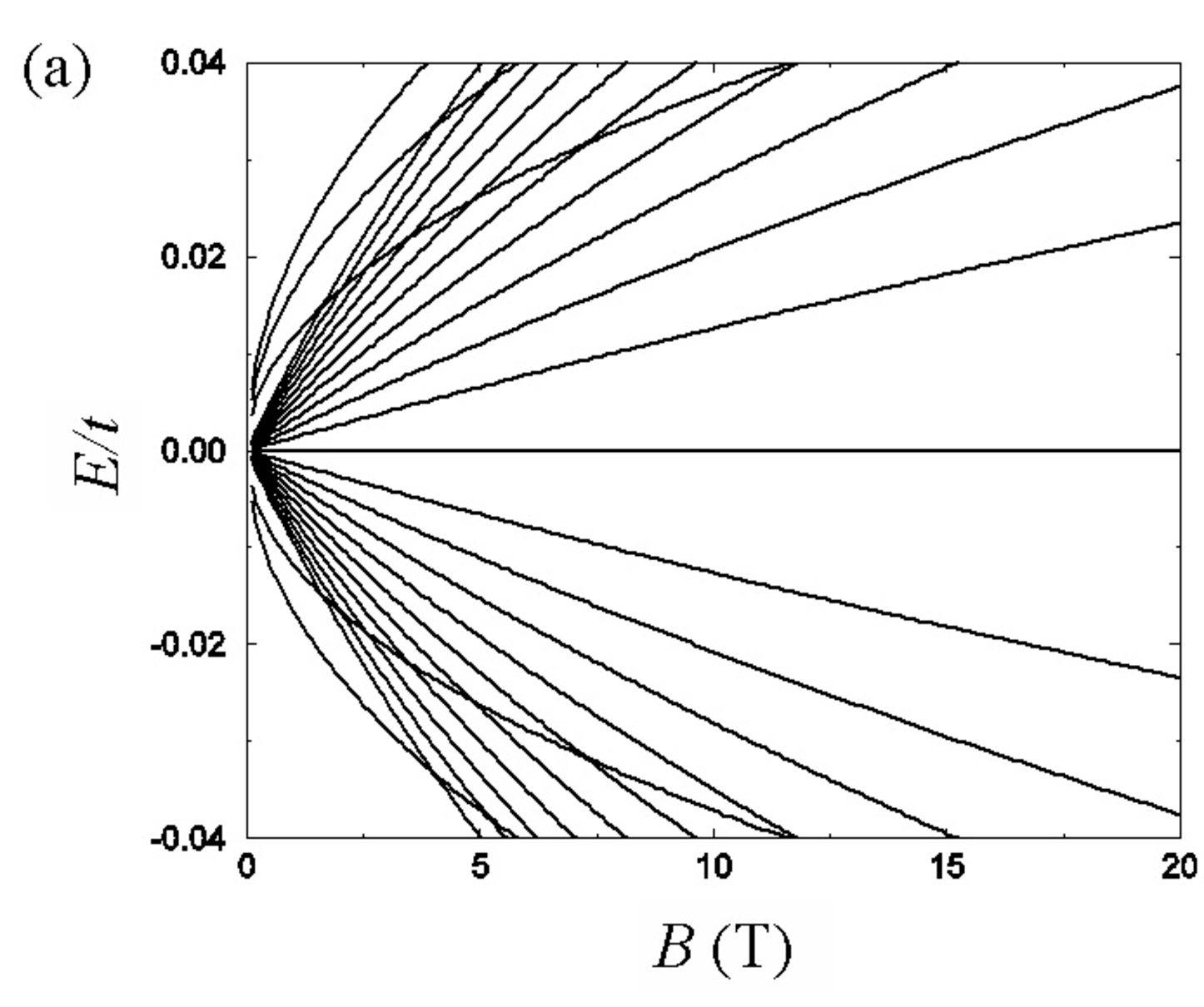}
\includegraphics[width=0.48\linewidth]{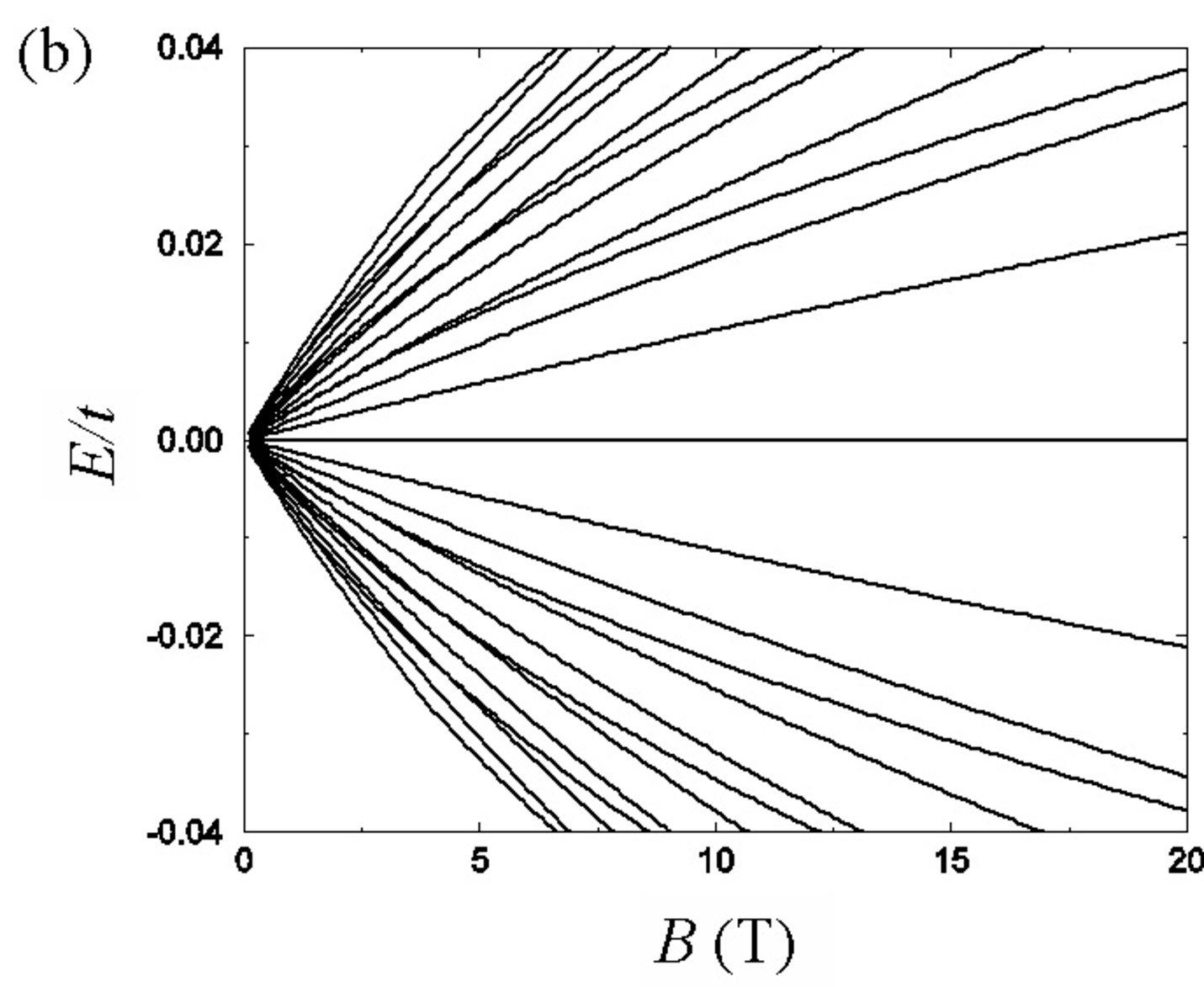}
\caption{Landau levels of (a) trilayer and (b) tetralayer graphene 
with AB stacking for nearest intralayer neighbor hopping $t=3$ eV and nearest interlayer neighbor hopping $t_{\perp}=0.1t$. Landau levels up to $n=10$
are shown.}
\label{fig:LL_AB}
\end{figure}
Figure \ref{fig:LL_AB} shows the Landau levels of AB stacked trilayer and
tetralayer graphene as a function of magnetic fields.
Note that the linear $B$ dependence expected for massive modes applies 
over a more limited field range when the mass is small.
For the trilayer, Landau levels are composed of massless Dirac spectra ($\propto B^{1/2}$)
and massive Dirac spectra ($\propto B$), while for the tetralayer,
Landau levels are all massive Dirac spectra.
This is consistent with the band structure analysis shown in Fig.~\ref{fig:band_AB}.

Note that the massive modes in Eq.~(\ref{eq:LL_massive}) have two zero-energy Landau levels for $n=-1$ and $0$, 
whereas the massless mode in Eq.~(\ref{eq:LL_massless}) has one for $n=0$.
There are therefore $N$ zero-energy Landau levels in both even and odd $N$ AB stacks.  
This property can also be understood directly from the 
Hamiltonian in Eq.~(\ref{eq:hamiltonian_AB_LL}), by eliminating 
negative $n$ basis states and rearranging rows to block diagonalize
the matrix.

\subsubsection{ABC stacking}
In the case of ABC stacking, a proper choice of the $n$-th Landau level basis at $K$ is\\
$(\alpha_{1,n-1},\beta_{1,n},\alpha_{2,n},\beta_{2,n+1},\alpha_{3,n+1},\beta_{3,n+2},\cdots)\,$
such that all the interlayer hopping terms are contained in the $n$-th Landau level Hamiltonian.
Then Eq.~(\ref{eq:hamiltonian_ABC}) reduces to
\begin{equation}
H_{\rm ABC}(n)=\left(
\begin{array}{ccccccc}
0              &\varepsilon_n     &0              &0              &0              &0              &               \\
\varepsilon_n     &0              &t_{\perp}      &0              &0              &0              &               \\
0              &t_{\perp}      &0              &\varepsilon_{n+1} &0              &0              &               \\
0              &0              &\varepsilon_{n+1} &0              &t_{\perp}      &0              &\cdots         \\
0              &0              &0              &t_{\perp}      &0              &\varepsilon_{n+2} &               \\
0              &0              &0              &0              &\varepsilon_{n+2} &0              &               \\
               &               &               &\cdots         &               &               &               \\
\end{array}
\right),
\end{equation}
where $\varepsilon_n=\sqrt{2 n}\hbar v /l$.

The low-energy spectrum can be obtained from the effective Hamiltonian in Eq.~(\ref{eq:hamiltonian_ABC_eff}).
For $n>0$, Landau levels are given by
\begin{eqnarray}
\varepsilon^{\pm}_{n}&=&\pm\hbar \omega_N \sqrt{n(n+1)\cdots (n+N-1)},
\end{eqnarray}
where $\hbar \omega_N=t_{\perp}(\sqrt{2}\hbar v / t_{\perp} l)^N\propto B^{N/2}$,
while for $n=-N+1,-N+2,\cdots,0$ they are zero.
Note that there are $N$ zero-energy Landau levels for ABC stacked $N$-layer graphene.

\begin{figure}[h]
\includegraphics[width=0.48\linewidth]{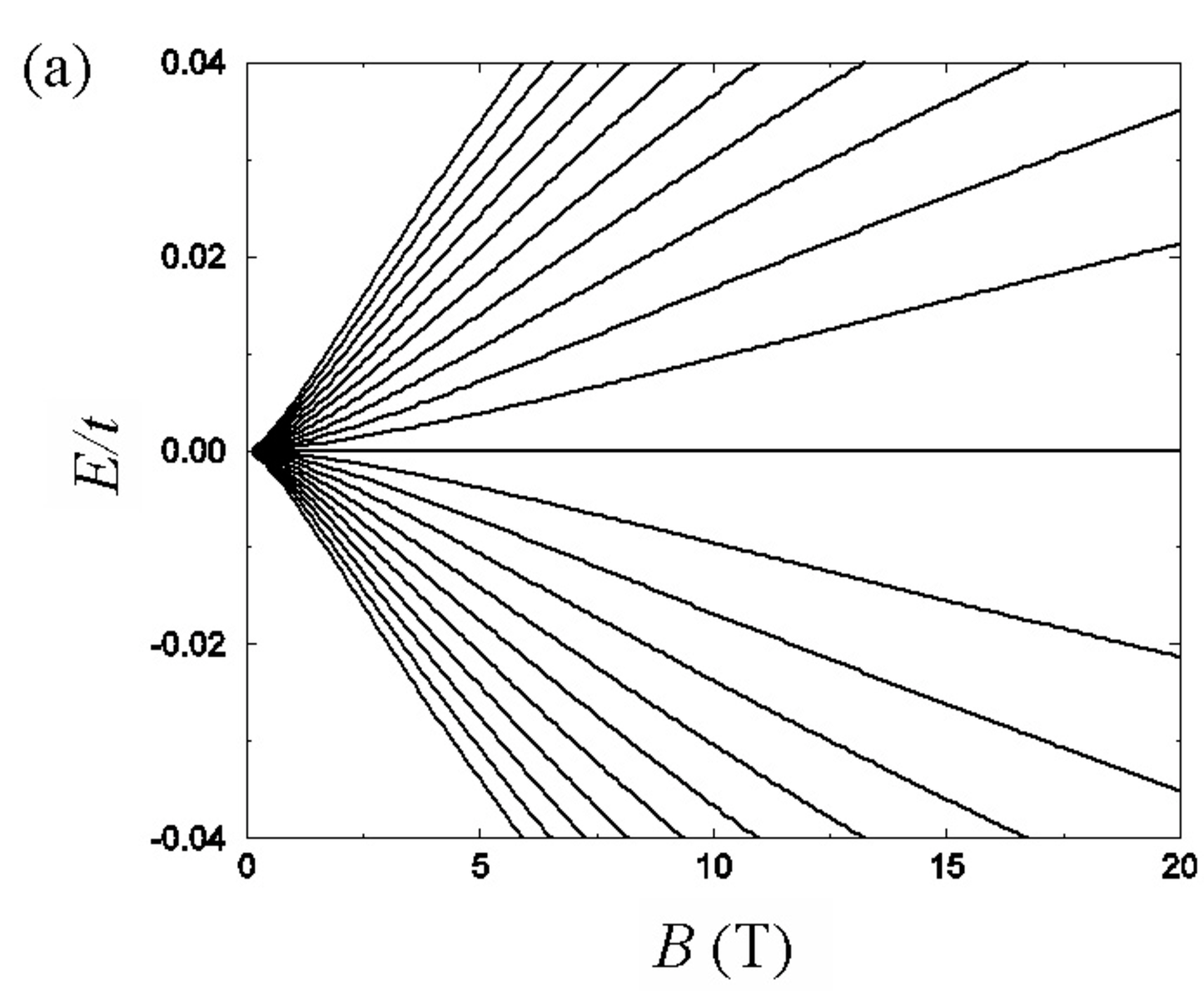}
\includegraphics[width=0.48\linewidth]{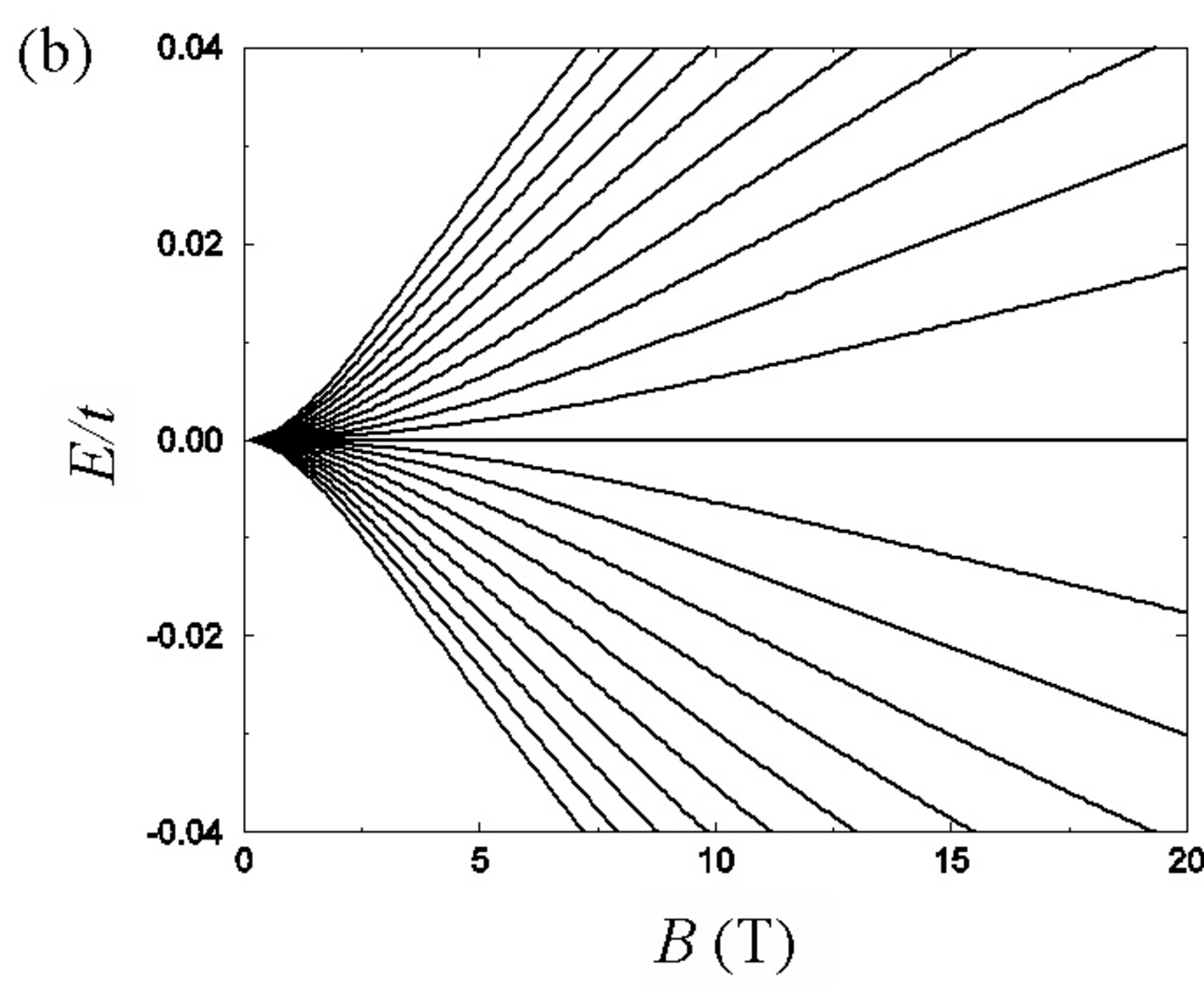}
\caption{Landau levels of (a) trilayer and (b) tetralayer graphene 
with ABC stacking for nearest intralayer neighbor hopping $t=3$ eV and nearest interlayer neighbor hopping $t_{\perp}=0.1t$. Landau levels
up to $n=10$ are shown.}
\label{fig:LL_ABC}
\end{figure}
Figure \ref{fig:LL_ABC} shows the Landau levels of ABC stacked trilayer and
tetralayer graphene as a function of magnetic fields.
For the trilayer, Landau levels are proportional to $B^{3/2}$, while for the tetralayer,
Landau levels are proportional to $B^2$.

\subsubsection{Arbitrary stacking}
It is straightforward to generalize the previous discussion to construct the Hamiltonian in Landau level basis for an arbitrarily stacked multilayer graphene system.
As discussed earlier, special care should be given for states with a negative index.
\begin{figure}[h]
\includegraphics[width=0.48\linewidth]{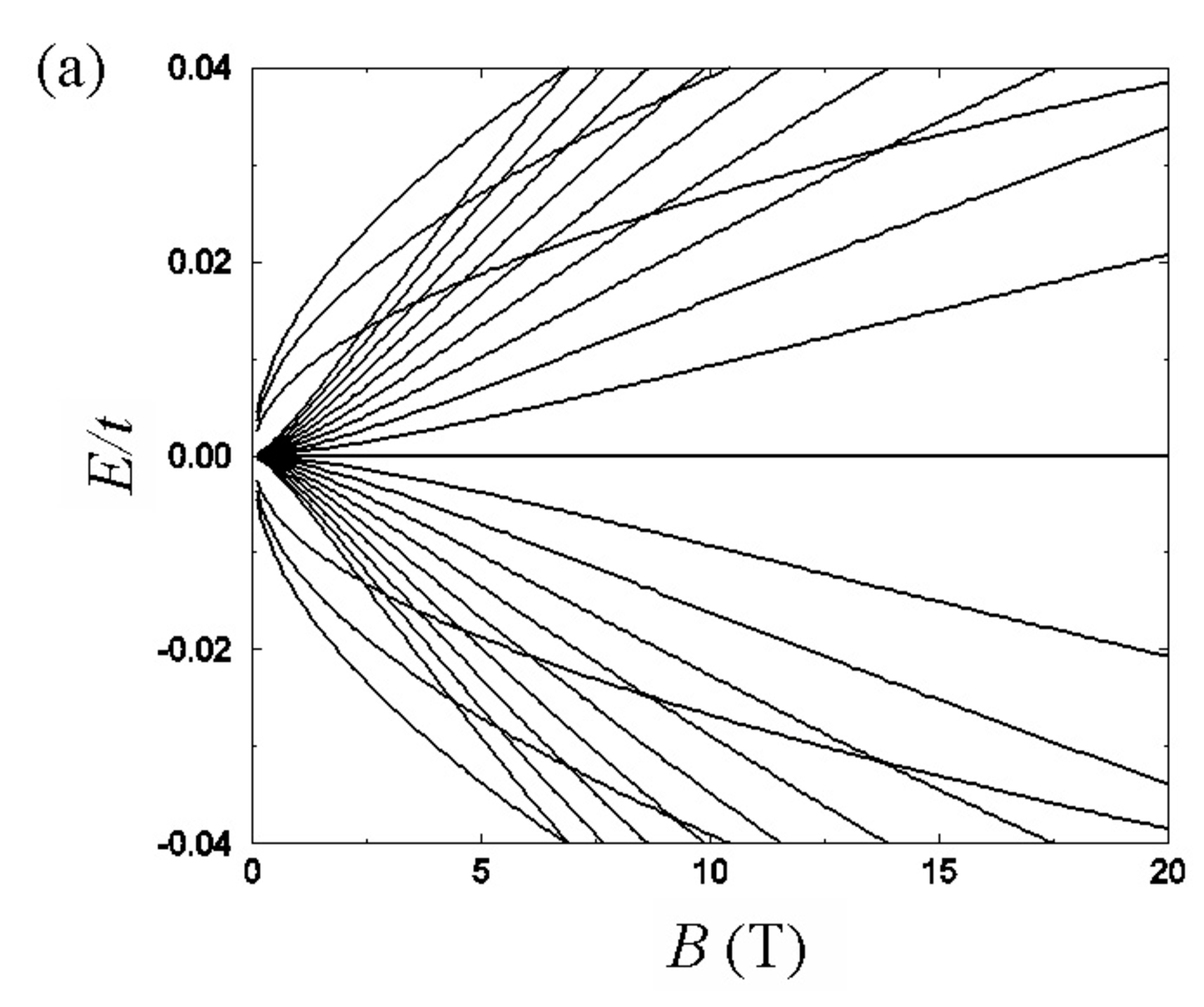}
\includegraphics[width=0.48\linewidth]{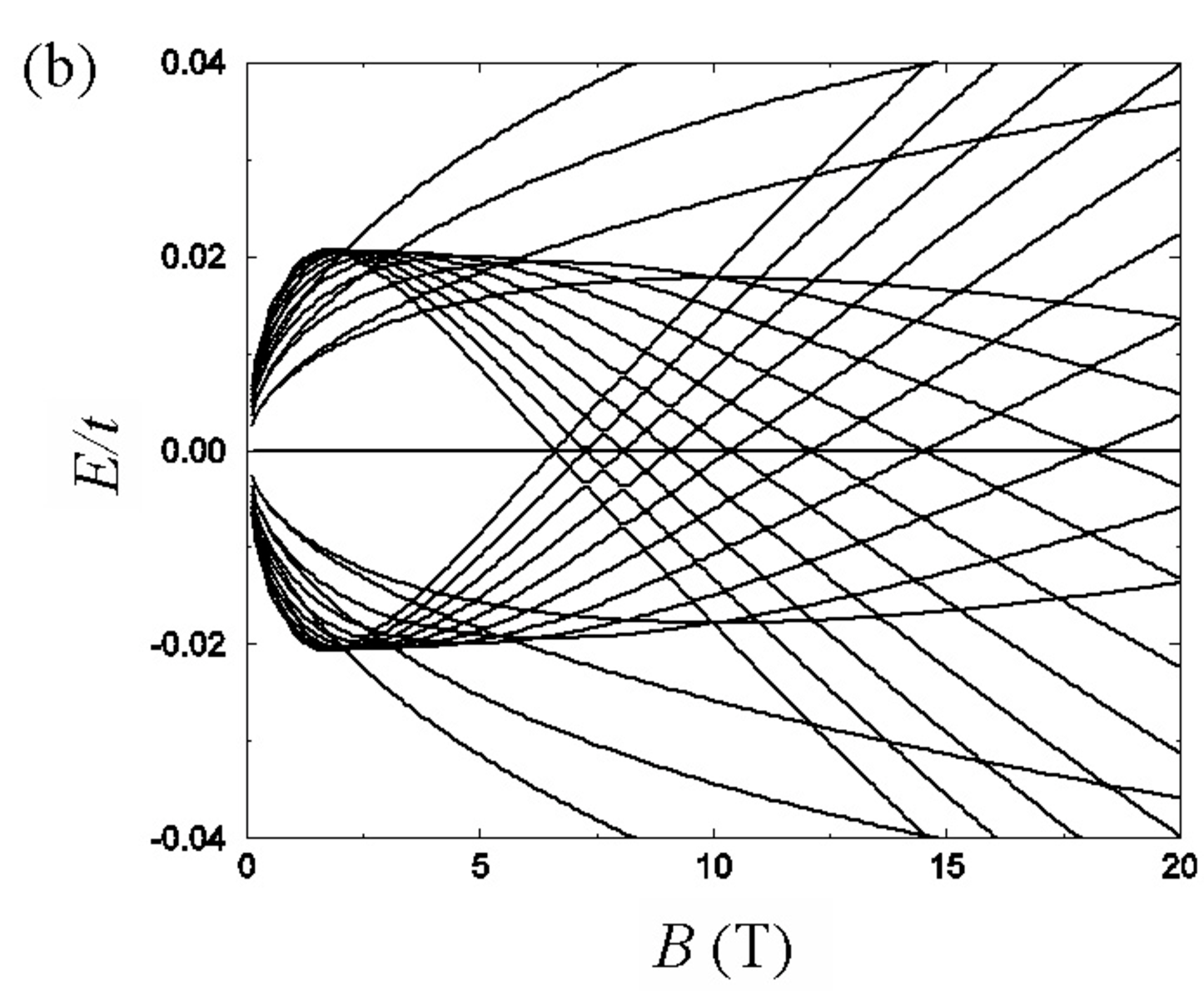}
\caption{Landau levels of tetralayer graphene with (a) ABCB stacking
and (b) ABBC stacking for nearest intralayer neighbor hopping $t=3$ eV and nearest interlayer neighbor hopping $t_{\perp}=0.1t$. Landau levels 
up to $n=10$ are shown.}
\label{fig:LL_arbitrary}
\end{figure}
Figure \ref{fig:LL_arbitrary} shows Landau levels of ABCB stacked tetralayer graphene and ABBC stacked tetralayer graphene.
For the ABCB stacked tetralayer graphene, the Landau levels look like a superposition
of $B^{1/2}$ and $B^{3/2}$ levels, which is consistent with Fig.~\ref{fig:band_arbitrary}(a).
For the ABBA stacked tetralayer graphene, 
there are Landau levels crossing the zero-energy line, which is consistent with Fig.~\ref{fig:band_arbitrary}(b).
Detailed low-energy Landau-level spectrum analysis will be presented in \S\ref{sec:chiral_decomposition}.

\subsection{Quantum Hall conductivity}
Applying the Kubo formula to a disorder-free systems we find that the conductivity tensor with an external magnetic field along $z$ is given by 
\begin{equation}
\sigma_{ij}(\omega)=-{e^2\over 2\pi\hbar l_B^2} \sum_n f_n \Omega_{ij}^n(\omega),
\end{equation}
where $f_n$ is Fermi factor of $n$-th energy state, $i,j=x,y$ and
\begin{equation}
\Omega_{ij}^n(\omega)=i \sum_{m\neq n}
\left[ 
{\left<n\right|\hbar v_i\left|m\right>\left<m\right|\hbar v_j\left|n\right>\over (\varepsilon_n-\varepsilon_m)(\varepsilon_n-\varepsilon_m+\hbar\omega+i\eta)}
-{\left<m\right|\hbar v_i\left|n\right>\left<n\right|\hbar v_j\left|m\right>\over (\varepsilon_n-\varepsilon_m)(\varepsilon_n-\varepsilon_m-\hbar\omega-i\eta)}
\right].
\end{equation}
Here $v_i$ is a velocity operator obtained by taking a derivative of the Hamiltonian $H({\bm p})$
with respect to $p_i$. Note that in case of multilayer graphene, the velocity operator
is constant, i.e. it does not depend on the Landau level index.

The appropriate quantized Hall conductivity is obtained by evaluating $\sigma_H=\sigma_{xy}(0)$.
Detailed analysis of the quantum Hall conductivity will be presented in \S\ref{sec:chiral_decomposition}.


\section{Chiral decomposition of energy spectrum}

\label{sec:chiral_decomposition}
In this section\footnote{The content of this section provides a more 
complete explanation of the arguments presented earlier in Ref.~10).}  
we demonstrate an unanticipated low-energy property of\linebreak 
graphene multilayers, which follows from an interplay between interlayer tunneling and the chiral
properties of low-energy quasiparticles in an isolated graphene sheet.  
Our conclusions apply in the strongest form to models with only
nearest-neighbor interlayer tunneling, but are valid over a broad field range 
as we explain below.  We find that the low-energy band structure of any graphene multilayer
consists of a set of independent pseudospin doublets.  Within each doublet,
the bands are described by a pseudospin Hamiltonian of the form
\begin{equation}
H_J({\bm k}) \, \propto  k^J \; [\, \cos(J \phi_{\bm k}) \, \tau^x \, \pm \, \sin(J \phi_{\bm k}) \, \tau^y \,],
\label{eq:chiralband}
\end{equation}
where $\tau^{\alpha}$ is a Pauli matrix acting on the doublet pseudospin, ${\bm k}$ is an
envelope function momentum measured from either the $K$ or $K'$ corner of the
honeycomb lattice's Brillouin-zone\cite{geim2007a,geim2007b}, $k=|{\bm k}|$, and $\phi_{\bm k}$ is the
orientation of ${\bm k}$.  The $\pm$ sign in Eq.~(\ref{eq:chiralband}) assumes the opposite signs in
graphene's $K$ and $K'$ valleys.  Following the
earlier work on graphene bilayers\cite{mccann2006}, we refer to $J$ as the chirality
index of a doublet.  In the presence of a perpendicular magnetic field $B$, $H_J({\bm k})$ yields $J$ Landau levels at 
$E = 0$ and $E \ne 0$ levels with $|E| \propto B^{J/2}$.  Taking the twofold spin and valley degeneracies into account,
the number of independent zero-energy band eigenstates at the Dirac
point (${\bm k}=0$) is therefore $8 N_{D}$, where $N_D$ is the number of pseudospin doublets.
We find that, although $N_{D}$ depends on the details of the stacking sequence,
\begin{equation}
\sum_{i=1}^{N_{D}} \; J_i \; = N
\label{eq:sumrule}
\end{equation}
in an $N$-layer stack.
It follows from
Eq.~(\ref{eq:sumrule}) that the Hall conductivity of an
$N$-layer stack has strong integer quantum Hall effects with plateau conductivities,
\begin{equation}
\sigma_{xy} \, = \,\pm {4 e^2\over h} \,\left({N\over 2}+n\right),
\label{eq:iqhe}
\end{equation}
where $n$ is a non-negative integer. 

\subsection{Partitioning rules}
The low-energy band and the Landau level structure can be read off the stacking diagrams illustrated in Fig.~\ref{fig:min_diagrams} by partitioning a stack using the following rules, which are justified in the following section.

(i) Identify the longest nonoverlapping segments within which there
are no reversals of stacking sense.  When there is ambiguity in the selection of nonoverlapping segments,
choose the partitioning which incorporates the largest number of layers.  Each segment (including for
interior segments the end layers
at which reversals take place) defines a $J$-layer partition of the stack and may be associated with a chirality $J$ doublet.
  
(ii)  Iteratively partition the remaining segments of the stack into
smaller $J$ elements, excluding layers
contained within previously identified partitions, until all layers are exhausted.

The chirality decompositions which
follow from these rules are summarized in Table \ref{tab:decomposition}.
Note that this procedure can result in $J=1$ doublets associated with separated 
single layers which remain at the last step in the partitioning process.

In applying these rules, the simplest case is cyclic ABC stacking for which there are
no stacking sense reversals and therefore a single $J=N$ partition.
In the opposite limit, AB stacking, the stacking sense is reversed in every layer and the rules
imply $N/2$ partitions with $J=2$ for even $N$, and when $N$ is odd a remaining $J=1$ partition.  Between these two limits, a rich variety of qualitatively distinct low-energy behaviors occur.
For example, in the ABCB stacked tetralayer, ABC is identified as a $J=3$ doublet and the remaining B
layer gives a $J=1$ doublet.  The low-energy band structure and the Landau level structure
of this stack, as illustrated in Figs.~\ref{fig:band_arbitrary}(a) and \ref{fig:LL_arbitrary}(a), have
two sets of low-energy bands with $|E|\propto k, k^3$,
Landau levels with $|E|\propto B^{1/2},B^{3/2}$, and four zero-energy Landau levels
per spin and valley.  All these properties are predicted by the partitioning rules.
We have explicitly checked that the rules correctly reproduce the low-energy electronic structure
for all stacking sequences up to $N=7$.
Because each layer is a member of one and only one partition, the partitioning rules imply
the chirality sum rule in Eq.~(\ref{eq:sumrule}).

\begin{table}[h]
\caption{Chirality decomposition for $N=3,4,5,6$ layer stacks.}
\begin{center}
\begin{tabular}{p{0.22\textwidth} p{0.22\textwidth} | p{0.22\textwidth} p{0.22\textwidth}}
\hline\hline
stacking & chirality & stacking & chirality\\
\hline
ABC      & 3         & ABCABC   & 6        \\
ABA      & 2+1       & ABCABA   & 5+1      \\
         &           & ABCACA   & 4+2      \\
ABCA     & 4         & ABCACB   & 4+2      \\
ABCB     & 3+1       & ABCBCA   & 3+3      \\
ABAB     & 2+2       & ABCBCB   & 3+2+1    \\
ABAC     & 1+3       & ABCBAB   & 3+2+1    \\
         &           & ABCBAC   & 3+3      \\
ABCAB    & 5         & ABABCA   & 2+4      \\
ABCAC    & 4+1       & ABABCB   & 2+3+1    \\
ABCBC    & 3+2       & ABABAB   & 2+2+2    \\
ABCBA    & 3+2       & ABABAC   & 2+1+3    \\
ABABC    & 2+3       & ABACAB   & 2+1+3    \\
ABABA    & 2+2+1     & ABACAC   & 1+3+2    \\
ABACA    & 1+3+1     & ABACBC   & 1+4+1    \\
ABACB    & 1+4       & ABACBA   & 1+5      \\
\hline\hline 
\end{tabular}
\label{tab:decomposition}
\end{center}
\end{table}

\subsection{Degenerate state perturbation theory}
We start from the well-known $J=1$ massless
Dirac equation\cite{geim2007a,geim2007b} ${\bm k}\cdot{\bm p}$ model for isolated sheets,
\begin{equation}
\label{eq:hamiltonian_MD}
H_{MD}({\bm p})=-\left(
\begin{array}{cc}
0              &v\pi^{\dagger}   \\
v\pi           &0                \\
\end{array}
\right),
\end{equation}
where $\pi = p_x + i p_y$ and $v$ is the quasiparticle velocity.
In the presence of an external magnetic field,
$\pi$ and $\pi^{\dagger}$ are proportional to the Landau level raising and lowering operators, so that Eq.~(\ref{eq:hamiltonian_MD})
implies the presence of one macroscopically degenerate Landau level at the Dirac point for each spin and valley, and therefore,
to the $N=1$ quantum Hall effect\cite{novoselov2005,zhang2005} of Eq.~(\ref{eq:iqhe}).
An $N$-layer stack has a two-dimensional band structure with $2N$ atoms per unit cell.
The Hamiltonian can be written as 
\begin{equation}
H = H_{\perp} + H_{\parallel},
\end{equation}
where $H_{\perp}$ accounts for interlayer tunneling and $H_{\parallel}$ for intralayer tunneling.  $H_{\parallel}$ is
the direct product of massless Dirac model Hamiltonians $H_{MD}$ for the sublattice pseudospin degrees of freedom of each layer.  We construct a low-energy Hamiltonian by first identifying the zero-energy eigenstates of $H_{\perp}$
and then treating $H_{\parallel}$ as a perturbation.

Referring to Fig.~\ref{fig:min_diagrams}, we see that
$H_{\perp}$ is the direct product of a set of finite-length 1D tight-binding chains, as shown in Eq.~(\ref{eq:chain}),
and a null matrix with dimension equal to the
number of isolated sites. 
The set of zero-energy eigenstates of $H_{\perp}$ consists of the states
localized on isolated sites and the single zero-energy eigenstates of each odd-length chain.

The low-energy effective Hamiltonian is evaluated by applying leading order
degenerate state perturbation theory to the zero-energy subspace.
The matrix element of the effective Hamiltonian between degenerate zero-energy states $r$ and $r'$
is given by\cite{sakurai1994}
\begin{equation}
\label{eq:perturbation}
\left<\Psi_r|H|\Psi_{r'}\right>= \left<\Psi_r\right| H_{\parallel}
\left[ \hat{Q}(-H_{\perp}^{-1}) \hat{Q} H_{\parallel} \right]^{n-1}\left|\Psi_{r'}\right>,
\end{equation}
where $n$ is the smallest positive integer for which the matrix element is nonzero,
and $\hat{Q}=1-\hat{P}$, $\hat{P}$ is a projection operator onto the zero-energy subspace.  To understand the
structure of this Hamiltonian, it is helpful to start with some simple examples.

\subsubsection{ABC stacking}

For ABC stacked $N$-layer graphene, the zero-energy states are the two isolated site states in bottom and top layers, $\alpha_1$ and $\beta_N$. $N-1$ sets of two-site chains form high-energy states.
Because $H_{\parallel}$ is diagonal in layer index and
$H_{\perp}$ (and hence $H_{\perp}^{-1}$) can change the layer index by one unit,
the lowest order at which $\alpha_1$ and $\beta_N$ are coupled is $n=N$.

According to Eq.~(\ref{eq:chain}), the wavefunction of each two-site chain is given by
\begin{equation}
\left|\Phi_{\sigma_r}\right>={1\over \sqrt{2}}\big(\left|\beta_r\right>+\sigma_r \left|\alpha_{r+1}\right>\big),
\end{equation}
with the energy $\epsilon_r=t_{\perp}\sigma_r$, where $\sigma_r=\pm 1$ and $r=1,2,\cdots,N-1$.  From Eq.~(\ref{eq:perturbation}),
\begin{eqnarray}
\label{eq:ABC_eff}
\left<\alpha_1|H|\beta_N\right>
&=&\left<\alpha_1\right| H_{\parallel}\left[\hat{Q}(-H_{\perp}^{-1}) \hat{Q} H_{\parallel}\right]^{N-1}\left|\beta_N\right> \nonumber \\
&=&\sum_{\{\sigma_r\}}{\left<\alpha_1|H_{\parallel}|\Phi_{\sigma_1}\right>\cdots\left<\Phi_{\sigma_{N-1}}|H_{\parallel}|\beta_N\right> \over (-\varepsilon_1)\cdots (-\varepsilon_{N-1}) } \nonumber \\
&=&-t_{\perp}\sum_{\{\sigma_r\}}{(-\sigma_1/2)\cdots(-\sigma_{N-1}/2)\over (-\sigma_1)\cdots (-\sigma_{N-1})} (\nu^{\dagger})^N \nonumber \\
&=&-t_{\perp}(\nu^{\dagger})^N \sum_{\sigma_1,\cdots,\sigma_{N-1}}{1\over 2^{N-1}} \nonumber \\
&=& \; -t_{\perp}(\nu^{\dagger})^N,
\end{eqnarray}
where $\nu = v\pi/t_{\perp}$. Here
$\left<\alpha_1|V|\Phi_{\sigma_1}\right>=-(1/\sqrt{2})t_{\perp}\nu^{\dagger}$,
$\left<\Phi_{\sigma_{N-1}}|V|\beta_N\right> =-(\sigma_{N-1}\linebreak /\sqrt{2})t_{\perp}\nu^{\dagger}$ and
$\left<\Phi_{\sigma_r}|V|\Phi_{\sigma_{r+1}}\right>=-(\sigma_r/2)t_{\perp}\nu^{\dagger}$ were used.
Thus, the effective Hamiltonian of $N$-layer graphene with ABC stacking has a
single $J=N$ doublet given by
\begin{equation}
H_N^{eff}=-t_{\perp} \left(
\begin{array}{cc}
0       & (\nu^{\dagger})^N \\
(\nu)^N & 0                \\
\end{array}
\right).
\end{equation}

\subsubsection{AB stacking}
For AB stacked $N$-layer graphene, the high-energy Hilbert space
consists of a single $N$-site 1D chain, excluding its zero-energy eigenstate when $N$ is odd.  
There is an isolated site in each layer which is connected to both its neighbors at order
$n=2$ forming an isolated site chain.  When $N$ is even, this chain
is diagonalized by $N/2$, $J=2$ doublets formed between $\alpha$-sublattice and $\beta$-sublattice
chain states\cite{guinea2006,koshino2007,manes2007,nakamura2008}. When $N$ is odd, the zero-energy chain state is mapped to
an equal-magnitude oscillating-sign linear combination of isolated site states by intralayer tunneling
at order $n=1$, yielding a $J=1$ doublet. The $(N-1)/2$, $J=2$ doublets are then
formed between $\alpha$-sublattice and $\beta$-sublattice isolated site chain states
in the orthogonal portion of the isolated state subspace.

Let us consider the low-energy spectrum of AB stacking in more detail. From Eq.~(\ref{eq:chain}) energy spectra and
wavefunctions of the single $N$-site chain are given by
\begin{eqnarray}
\label{eq:chain_AB}
\varepsilon_r&=&2 t_{\perp} \cos\theta_r ,\nonumber\\
\left|\Phi_r\right>&=&\sqrt{2\over N+1}\big(\sin\theta_r\left|\beta_1\right>+\sin 2\theta_r\left|\alpha_2\right> +\sin 3\theta_r\left|\beta_3\right>+\sin 4\theta_r\left|\alpha_4\right>\cdots\big), \qquad
\end{eqnarray}
where $\theta_r={r\pi \over N+1}$ and $r=1,2,\cdots,N$.

First, let us consider the case with even $N$.
Then the low-energy states come from the isolated sites or equivalently their superpositions.
Let us define
\begin{equation}
\left|\Psi_r\right>=\sqrt{2\over N+1}\big(\sin\theta_r e^{-i\phi}\left|\alpha_1\right>+\sin 2\theta_re^{i\phi}\left|\beta_2\right> 
+\sin 3\theta_re^{-i\phi}\left|\alpha_3\right>+\sin 4\theta_re^{i\phi}\left|\beta_4\right>\cdots\big)
\end{equation}
such that
\begin{equation}
\left<\Psi_r|V|\Phi_s\right>=-\delta_{r,s}|\nu|,
\end{equation}
where $e^{i\phi}=\nu/|\nu|$.
Then the matrix elements between the low-energy states are given by the second order perturbation theory:
\begin{equation}
\left<\Psi_r|H|\Psi_{r'}\right>=\sum_{s=1}^{N} {\left<\Psi_r|V|\Phi_s\right>\left<\Phi_s|V|\Psi_{r'}\right>\over (-\varepsilon_s)} 
=-\delta_{r,r'} (t_{\perp}^2/\varepsilon_r)|\nu|^2.
\end{equation}
Note that $\varepsilon_r=-\varepsilon_{N+1-r}$ and these two modes
form a 2-chiral system with energies $\pm |\varepsilon_r|$.
The chirality can be manifested clearly if we define
\begin{eqnarray}
\left|\tilde{\alpha}_r\right>&=&{e^{i\phi}\over
\sqrt{2}}\left(\left|\Psi_r\right>+\left|\Psi_{N+1-r}\right>\right),
\nonumber \\
\left|\tilde{\beta}_r\right>&=&{e^{-i\phi}\over \sqrt{2}}\left(\left|\Psi_r\right>-\left|\Psi_{N+1-r}\right>\right). 
\end{eqnarray}
Then the Hamiltonian of the 2-chiral system for $r=1,2,\cdots,N/2$ is given by
\begin{equation}
\label{eq:chiral2}
H_r=-{t_{\perp}^2\over \varepsilon_r}\left(
\begin{array}{cc}
0          & (\nu^{\dagger})^2 \\
(\nu)^2 & 0                    \\
\end{array}
\right)
=-\left(
\begin{array}{cc}
0                  & (\pi^{\dagger})^2\over 2 m_r \\
(\pi)^2\over 2 m_r & 0                            \\
\end{array}
\right)
\end{equation}
in a $(\tilde{\alpha}_r,\tilde{\beta}_r)$ basis with $m_r v^2=t_{\perp}\cos\left({r\pi\over N+1}\right)$.
Thus the system is described by a combination of massive Dirac modes with different masses.

For odd $N$, there is a zero-energy state in the $N$-site chain at $r=(N+1)/2$ in Eq.~(\ref{eq:chain_AB}).
Thus in addition to the massive modes, there exists one massless Dirac mode,
\begin{equation}
\left<\Psi_{N+1\over 2}|V|\Phi_{N+1\over 2}\right>=-|\nu|.
\end{equation}
Thus the system is described by one massless Dirac mode and a combination of massive Dirac modes with different masses.

\subsubsection{ABC+B type stacking}
A more complex and more typical example is realized by
placing a single reversed layer on top of ABC stacked $N$-layer graphene with $N>2$.
Note that the last chain has three sites, thus it has a zero-energy state $\beta_{N+1}^{-}$ defined by
\begin{equation}
\left|\beta_{N+1}^{-}\right>={1\over \sqrt{2}}\left(\left|\beta_{N+1}\right>-\left|\beta_{N-1}\right>\right),
\end{equation}
and two high-energy states with energies $\sqrt{2}\sigma_{N-1} t_{\perp}$ defined by
\begin{equation}
\left|\Phi_{\sigma_{N-1}}\right>={1\over 2}\left|\beta_{N-1}\right>+{\sigma_{N-1}\over \sqrt{2}}\left|\alpha_{N}\right>+{1\over 2}\left|\beta_{N+1}\right>,
\end{equation}
where $\sigma_{N-1}=\pm 1$.
Then the first-order perturbation theory gives
\begin{equation}
\left<\alpha_{N+1}|H|\beta_{N+1}^{-}\right>=-{t_{\perp}\over \sqrt{2}}\nu{\dagger},
\end{equation}
suggesting the existence of the massless Dirac mode with a {\em reduced} velocity.

Similarly as Eq.~(\ref{eq:ABC_eff}), we obtain
\begin{equation}
\label{eq:hamiltonian_ABCB_eff}
H_{N+1}^{eff}=-t_{\perp}\left(
\begin{array}{cccc}
0                  & {\nu^{\dagger}\over \sqrt{2}}          & 0                            & {(\nu^{\dagger})^2\over 2}  \\
{\nu\over\sqrt{2}} & 0                                      & -{(\nu)^{N-1}\over \sqrt{2}} & 0                           \\
0                  & -{(\nu^{\dagger})^{N-1}\over \sqrt{2}} & 0                            & {(\nu^{\dagger})^N\over 2}  \\
{\nu^2\over 2}     & 0                                      & {(\nu)^N\over 2}             & 0                           \\
\end{array}
\right),
\end{equation}
using a $(\alpha_{N+1},\beta_{N+1}^{-},\alpha_1,\beta_N)$ basis.
The first $2\times 2$ block in Eq.~(\ref{eq:hamiltonian_ABCB_eff}) gives a $J=1$ doublet with a reduced velocity.
Note that the matrix in Eq.~(\ref{eq:hamiltonian_ABCB_eff}) is not block diagonal thus we cannot simply say that the second $2\times 2$ matrix block is a $N$-chiral system. 
The $J=N$ doublet in this instance includes both the $(\alpha_1,\beta_N)$ subspace contribution and an equal contribution due to perturbative coupling to the $(\alpha_{N+1},\beta_{N+1}^{-})$ subspace. Using a similar perturbation theory shown in Eq.~(\ref{eq:effective}), we can obtain higher order correction by integrating out the massless Dirac mode which forms a higher energy state.
Then the final Hamiltonian is reduced to
\begin{equation}
\label{eq:hamiltonian_ABC+B_eff}
H_{N+1}^{eff}\approx H_1\otimes H_N,
\end{equation}
where
\begin{equation}
H_1=-t_{\perp}\left(
\begin{array}{cc}
0            & \nu^{\dagger}/\sqrt{2} \\
\nu/\sqrt{2} & 0                      \\
\end{array}
\right), \ \
H_N=-t_{\perp}\left(
\begin{array}{cc}
0       & (\nu^{\dagger})^N \\
(\nu)^N & 0                 \\
\end{array}
\right).
\end{equation}
This means that the combined system can be described by a combination of one $1$-chiral system with reduced velocity and one $N$-chiral system.
Note that stacking a layer with an opposite handedness partitions a system into systems with different chiralities.

\subsubsection{Arbitrary stacking}
The relationship between the electronic structure of a general stack and
the partitioning procedure explained above can be understood as follows.  

(i) First, note that a partition with chirality $J$ has isolated sites in its terminal layers that are coupled at order $J$ in perturbation theory.
In the case of $J=1$ partition, the chain opposite to the
single isolated site always has an odd length and provides the zero-energy partner;
isolated site to chain coupling therefore always occurs at first order.

(ii) Next, consider the perturbation theory, truncating at successively higher orders.
When truncated at first order, the $J=1$ partitions are isolated
by higher $J$ blocks within which the Hamiltonian vanishes.  Each $J=1$
partition therefore yields a separate massless Dirac equation
with velocities\footnote{The velocity of the $J=1$ doublets is determined by the strength of the coupling between the odd-length chain zero-energy state and isolated states on the sublattice opposite to the chain ends. For a chain of length $2N-1$, the chain's zero-energy state has nonzero amplitude on the $N$ odd-index sites. The velocity is reduced from the single sheet velocity by a factor of $\sqrt{M/N}$, where $M$ is the number of isolated sites opposite to the $N$ odd-index sites.  In a similar manner, higher $J$ doublet Hamiltonians are sometimes altered by a multiplicative factor by perturbative coupling to smaller $J$ doublets
as in the single reversed layer example.}
that can be smaller than the graphene sheet Dirac velocity.  
When the perturbation theory is truncated at second order,
the Hamiltonian becomes nonzero within the $J=2$ partitions.  The eigenenergies
within the $J=1$ partitions are parametrically larger, and the
Hamiltonian within the $J>2$ partitions is still zero.  To leading order therefore,
the $J=2$ partitions are separated, and their isolated states are coupled at
the second order in perturbation theory so that each provides a $J=2$ doublet such as that of an
isolated bilayer.  If two or more $J=2$ partitions are adjacent, then their
Hamiltonians do not separate.  In this case, there is a chain of second order
couplings between isolated states, such as those of an even-length AB stack, but
the end result is still $J=2$ doublet for each $J=2$ partition.

(iii) The identification between partitions and chiral doublets can be established by
continuing this consideration up to the highest values of $J$ which occur for a particular
stack.

(iv) Then, the effective Hamiltonian of any $N$-layer graphene is as follows: 
\begin{equation}
H_N^{eff}\approx H_{J_1} \otimes H_{J_2} \otimes \cdots \otimes H_{J_{N_D}},
\end{equation}
with the sum rule in Eq.~(\ref{eq:sumrule}).
Note that $N_D$ is half the sum of the number of isolated sites and the number of odd-length chains.

\subsection{Discussion}
\subsubsection{Effects of remote hopping}
The minimal model we have used to derive these results is approximately 
valid in the broad intermediate magnetic field $B$ range between $\sim 10$ and $\sim 100$ T,
over which the intralayer hopping energy in field ($\sim \hbar v/\ell$ where $\ell = \sqrt{\hbar c/e|B|} \sim 25\, {\rm nm}/[B({\rm T})]^{1/2}$ is the magnetic length) is larger than the distant neighbor interlayer 
hopping amplitudes that we have neglected ($\gamma_2 \sim -20 $ meV), 
but still smaller than $t_{\perp}$.
For example, if we consider $\alpha_1\rightarrow \alpha_3$ hopping process in ABA stacked trilayer in Fig.~\ref{fig:min_diagrams}, the valid range of magnetic field for the minimal model is given by 
\begin{equation}
|\gamma_2| < {(\hbar v/l)^2 \over t_{\perp}} < t_{\perp}.
\end{equation}

\begin{figure}[htb]
\begin{center}
\includegraphics[width=0.9\linewidth]{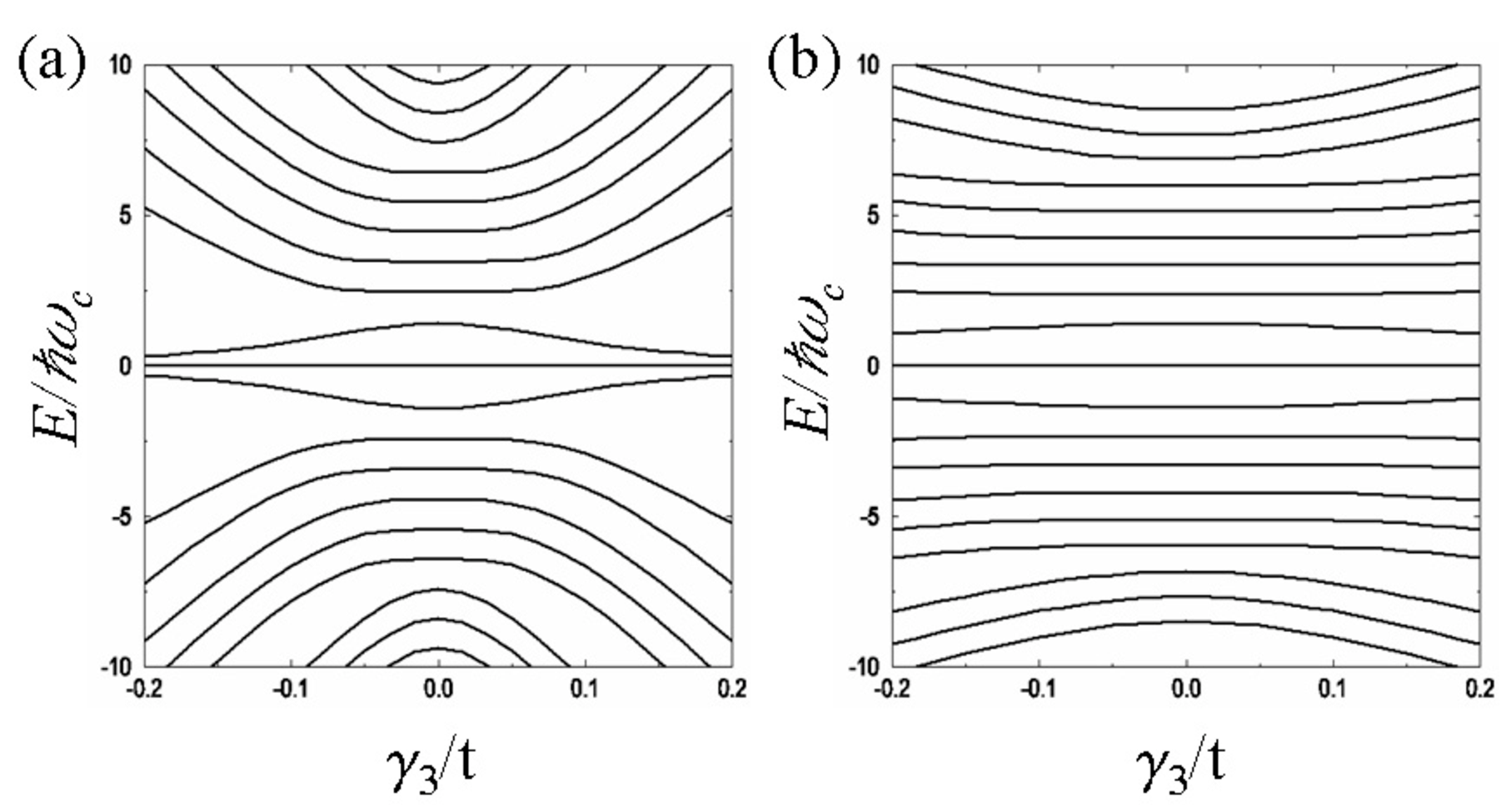}
\caption{Landau level spectrum near the $K$ valley as a function of $\gamma_3$ for an AB stacked bilayer for (a) $B=0.1$ T and (b) $B=1$ T. Here $t=3$ eV, $t_{\perp}=0.1t$, and $\omega_c=eB/mc$, with $m=t_{\perp}/2v^2$, were used.}
\label{fig:2NN_AB}
\end{center}
\end{figure}

\begin{figure}[htb]
\begin{center}
\includegraphics[width=0.9\linewidth]{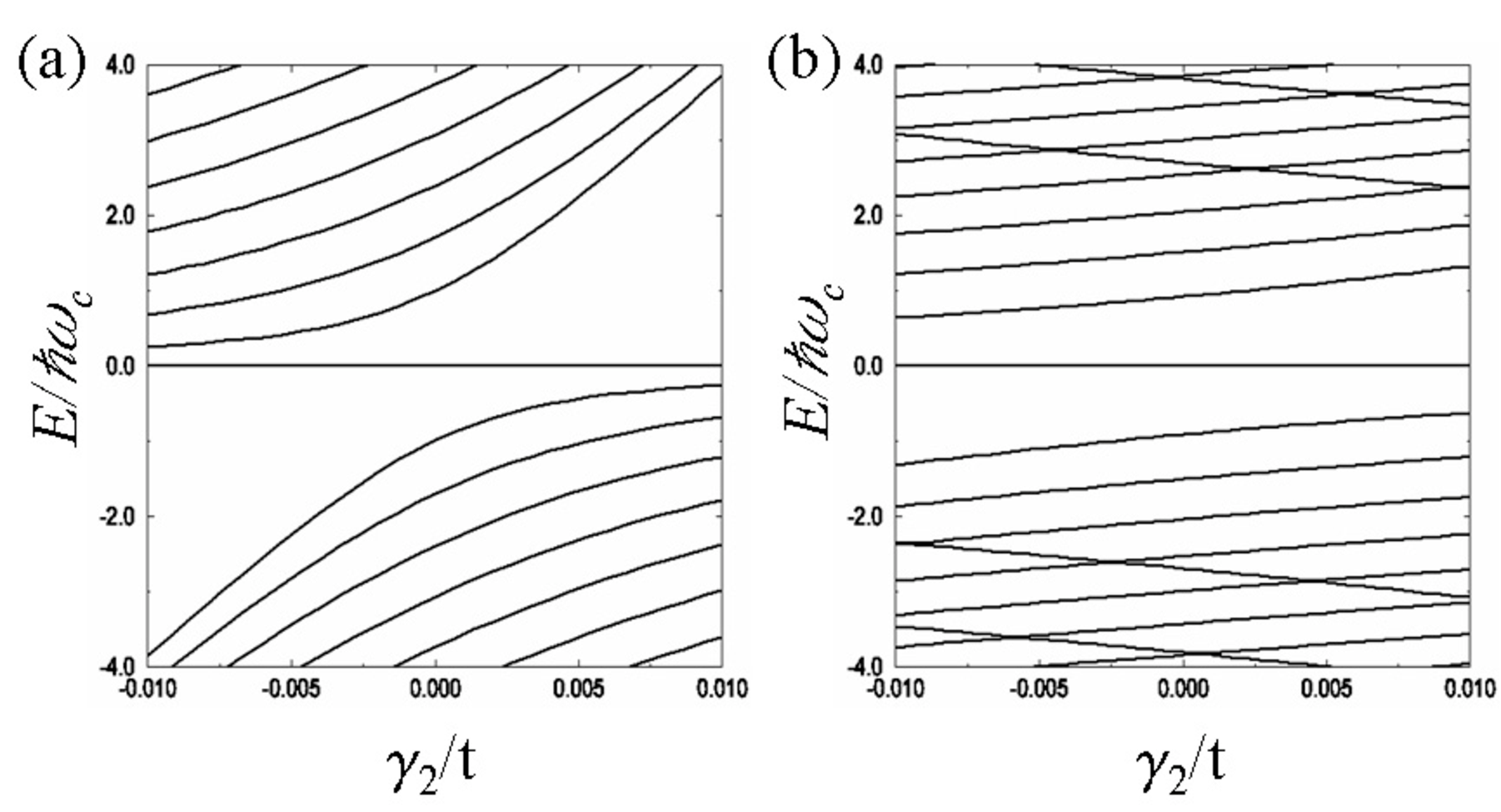}
\caption{Landau level spectrum near the $K$ valley as a function of $\gamma_2$ for an ABA stacked trilayer for (a) $B=1$ T and (b) $B=10$ T. Here $t=3$ eV, $t_{\perp}=0.1t$, and $\omega_c=eB/mc$, with $m=t_{\perp}/2v^2$, were used.  Note that for this case the 
Landau level structures near $K$ and $K'$ valleys are not identical.}
\label{fig:2NN_ABA}
\end{center}
\end{figure}

When $\gamma_2$ does not play an important role (in $N=2$ stacks, for example), the lower limit of the validity range 
is parametrically smaller.  The minimum field in bilayers has been estimated to be $\sim 1$ T\cite{mccann2006}, by comparing intralayer hopping with the $\gamma_3 \sim 0.3$ eV interlayer hopping amplitude,
\begin{equation}
\hbar v_3/l  < {(\hbar v/l)^2 \over t_{\perp}} < t_{\perp},
\end{equation}
where $v_3=(\sqrt{3}/2)a\gamma_3/\hbar$ and $a$ is a lattice constant of graphene.

Figures \ref{fig:2NN_AB} and \ref{fig:2NN_ABA} show the Landau level spectrum at the $K$ valley as a function of $\gamma_3$ for an AB stacked bilayer, and as a function of $\gamma_2$ for an ABA stacked trilayer, respectively. In the case of the bilayer, the dependence of the Landau levels on $\gamma_3$ is weak for $B$ larger than 1 T, whereas in the case of the trilayer, the Landau level spectrum still strongly depends on $\gamma_2$ for $B=1$ T, but the dependence becomes weak for $B$ above 10 T, confirming the above argument.

\subsubsection{Quantum Hall effect}
\begin{figure}[htb]
\begin{center}
\includegraphics[width=0.7\linewidth]{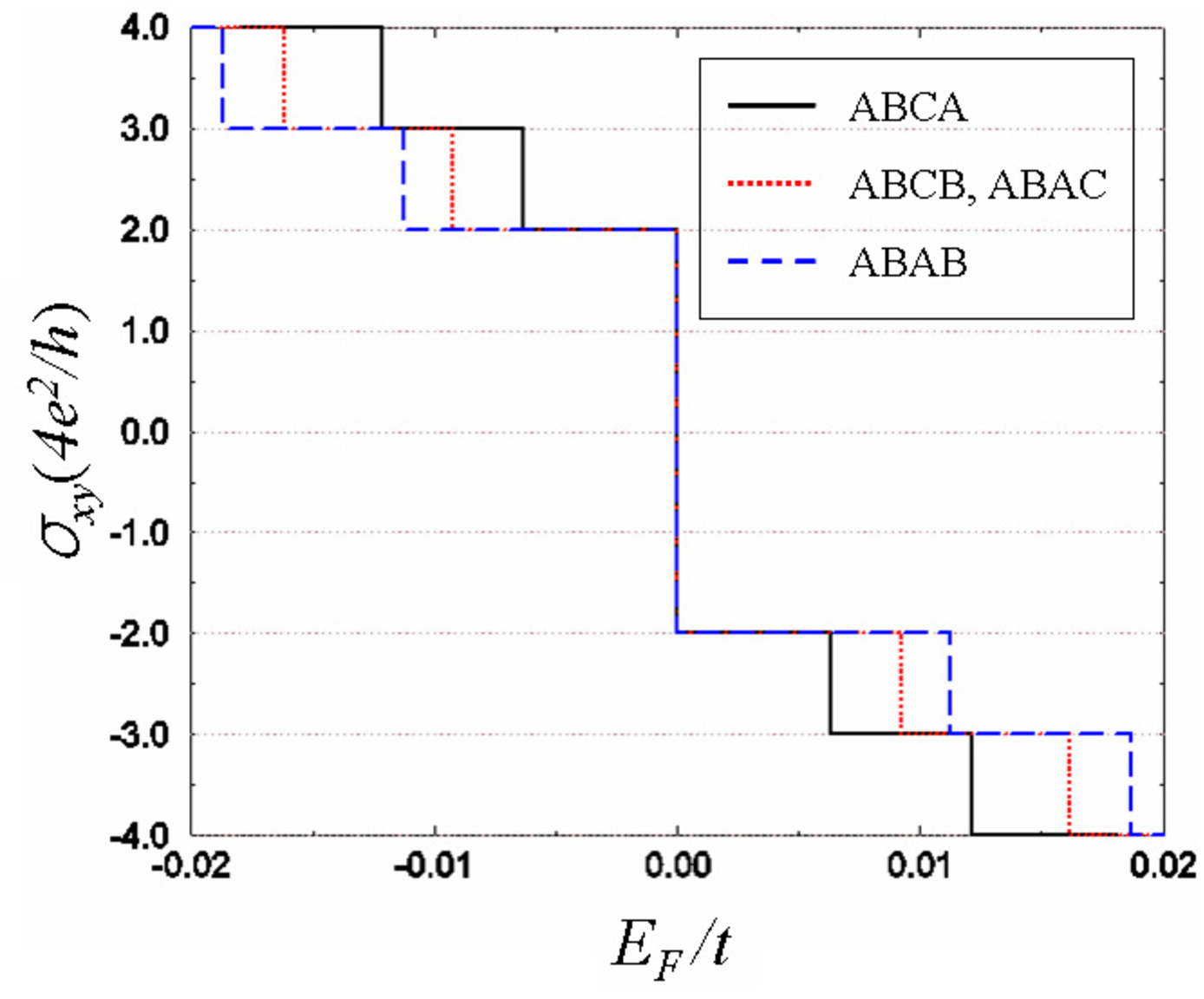}
\caption{(Color online) Noninteracting system Hall conductivity as a function of the Fermi energy
for all inequivalent four-layer graphene stacks when $B=10$ T, $t=3$ eV, and $t_{\perp}=0.1t$.
The dependence of the Hall conductivity on Fermi energy is simply related to the dependence of the Hall conductivity on total electron density.  
The Hall conductivity calculations shown in this figure assume neutralizing ionized donors spread equally between the four layers. 
}
\label{fig:IQHE4}
\end{center}
\end{figure}

In Fig.~\ref{fig:IQHE4}, we plot the noninteracting Hall conductivity as a
function of Fermi energy for the four distinct four-layer stacks.
When electron-electron interactions are included at an electrostatic mean-field (Hartree) level
and the neutralizing ionized dopants (responsible for the Fermi energy shift away from the
Dirac point) are assumed to be equally distributed among the layers,
the Landau levels with $E \ne 0$ are shifted by electrostatic potential differences between
the layers.  There is, however, no influence of electrostatics on the $E=0$ levels.  
This property follows from the perfect particle-hole symmetry of the models we employ,
which implies a uniform charge distribution among the layers at the neutrality point.
Remote ($\gamma_2$ 2nd neighbor) interlayer hopping {\em does} shift the 
$E = 0$ Landau level in the ABAB stacked tetralayer and weakly lifts the degeneracy responsible for the 
large jump between the $\pm (4e^2/h) N/2$ Hall plateaus.  This example demonstrates a 
tendency toward the grouping of $N$ spin and valley degenerate Landau levels 
very close to $E=0$ in general $N$-layer stacks even when remote neighbor hopping is 
included.  Small gaps between these Landau levels are unlikely to lead to 
Hall plateaus unless disorder is very weak.  When disorder is weak, on the other hand,
electron-electron interaction effects beyond Hartree level are likely to be important and lead to 
strong quantum Hall effects at many filling factors, often ones 
associated with broken symmetries of different types.\cite{abanin2006,nomura2006,alicea2006,goerbig2006,yang2006,barlas2008}. 
The property that the Hall conductivity will tend to jump by four units on crossing the Dirac point for arbitrarily stacked tetralayer 
graphene is the most obvious experimental manifestation of the chirality sum rule discussed in this paper.
In practice charged multilayers ($E_F\neq 0$) would normally be prepared by placing the system on one side of an electrode and gating.
Even though gating will redistribute charge and shift electric potentials differently in different layers, the Landau level bunching we discussed
should still be clearly reflected in quantum Hall effect measurements.

\subsubsection{Effects of the same stacking inside}
The analysis presented so far is based on the assumption that stacking one layer 
directly on top of its neighbor, AA stacking, is not allowed. 
When interior AA stacking does occur, we can still apply a similar diagram analysis
and identify the zero-energy states at the Dirac point. 
In this case, however, zero-energy states can appear not only at the Dirac points but also at 
other points in momentum space. The degenerate state perturbation theory at the Dirac point discussed so far
therefore cannot completely
capture the low-energy states.

\begin{figure}[t]
\begin{center}
\includegraphics[width=0.5\linewidth]{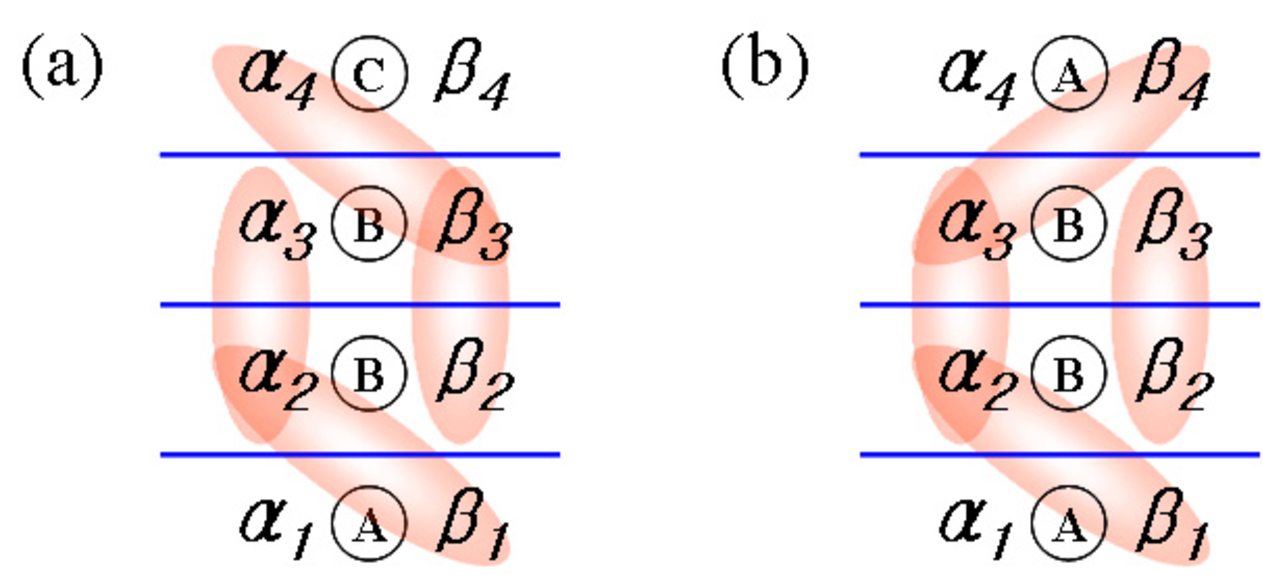}
\caption{(Color online) Stacking diagrams for tetralayer graphene with (a) ABBC stacking and (b) ABBA stacking. Shaded ovals link nearest interlayer neighbors.}
\label{fig:AA_inside}
\end{center}
\end{figure}

As an example, let us  consider ABBC stacked tetralayer graphene, as illustrated in Fig.~\ref{fig:AA_inside}(a).
Here, in addition to $\alpha_1$ and $\beta_4$, there are two zero-energy states at each three-site-chain defined by
\begin{eqnarray}
\left|\tilde{\beta}_1\right>&=&{1\over
\sqrt{2}}\left(\left|\beta_1\right>-\left|\alpha_3\right>\right),
\nonumber \\
\left|\tilde{\alpha}_4\right>&=&{1\over \sqrt{2}}\left(\left|\alpha_4\right>-\left|\beta_2\right>\right). 
\end{eqnarray}
Thus the matrix elements between low-energy states are given by
\begin{equation}
\left<\alpha_1|H|\tilde{\beta}_1\right>=\left<\tilde{\alpha}_4|H|\beta_4\right>=-{t_{\perp}\over \sqrt{2}}\nu^{\dagger}.
\end{equation}
Therefore the system can be described by two massless Dirac modes with reduced velocity,
as shown in Figs.~\ref{fig:band_arbitrary}(b) and \ref{fig:LL_arbitrary}(b).

Another example is ABBA stacked tetralayer graphene, as illustrated in Fig.~\ref{fig:AA_inside}.(b).
In this case, there are two zero-energy states at $\alpha_1$ and $\alpha_4$.
The high-energy states $\Phi_r$ are given by Eq.~(\ref{eq:chain}) with $N=4$, thus we get
\begin{equation}
\left<\alpha_1|H|\alpha_4\right>=\sum_{r=1}^{4} {\left<\alpha_1|V|\Phi_r\right>\left<\Phi_r|V|\alpha_4\right>\over (-\epsilon_r)} 
=-c t_{\perp}|\nu|^2,
\end{equation}
where $c={1\over 5}\sum_r \sin\left({r\pi\over 5}\right)\sin\left({4r\pi\over 5}\right)/\cos\left({r\pi\over 5}\right)=-1$.
Here the low-energy state is composed of one {\em non-chiral} massive mode.
Note that because of the non-chirality, there are no zero-energy Landau levels.

\subsubsection{Pseudospin magnetism}

\begin{figure}[htb]
\begin{center}
\includegraphics[width=0.9\linewidth]{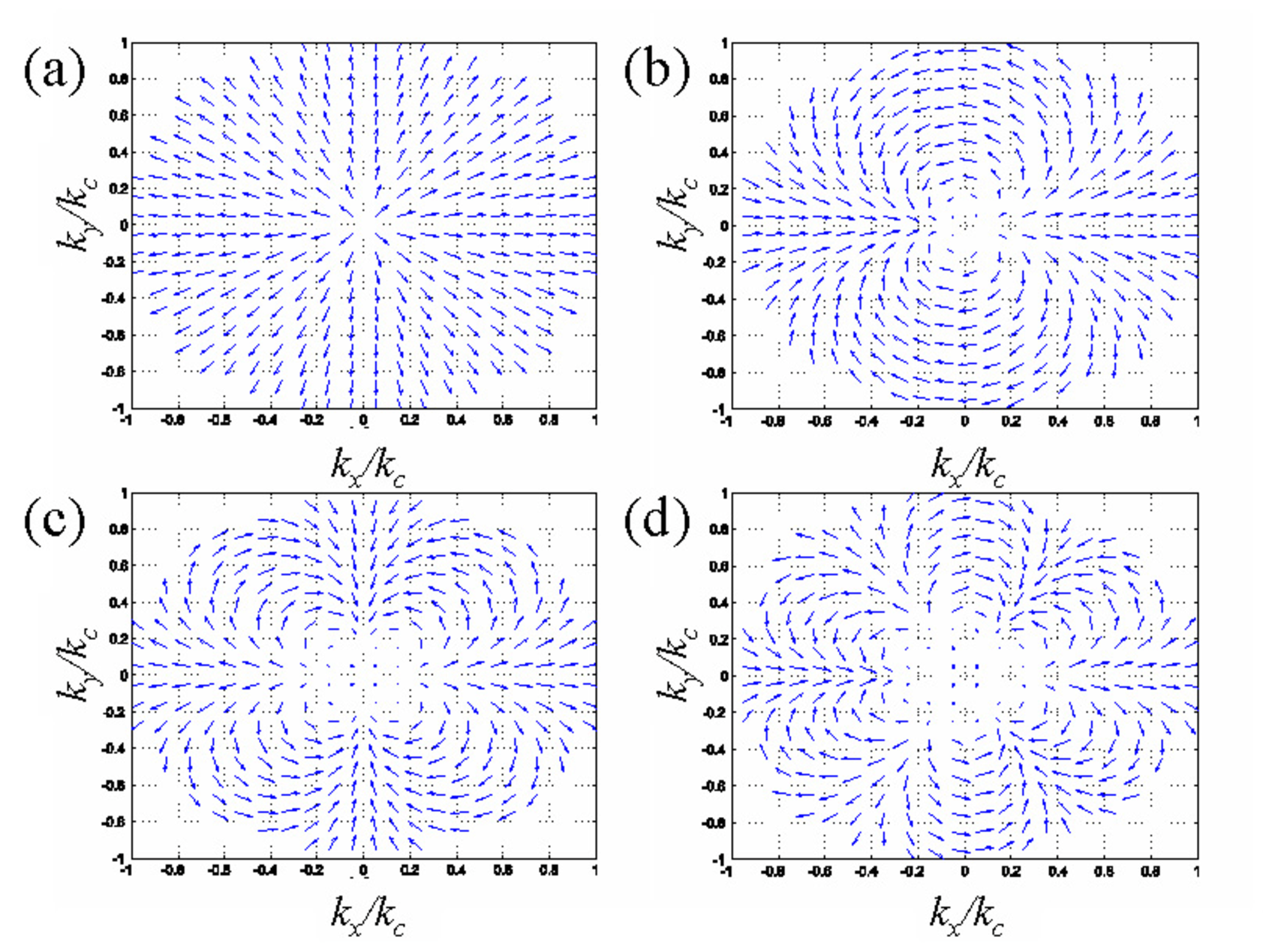}
\caption{(Color online) In-plane projected pseudospin orientation of (a) $J=1$, (b) $J=2$, (c) $J=3$ and (d) $J=4$ chiral 2D electron system for a neutral, unbiased system with coupling constant $\alpha\equiv e^2/\epsilon \hbar v =1$ where $\epsilon$ is the dielectric constant. For $J>1$, the arrows are shorter in the core of the momentum space vortex because the pseudospins in the core have rotated spontaneously toward $\hat{z}$ or $-\hat{z}$ direction indicating the pseudospin magnetic state.}
\label{fig:vortex}
\end{center}
\end{figure}


Finally, we note that in the presence of electron-electron interactions, chiral two-dimensional electron system (C2DES) tends toward momentum-space vortex states in which charge is spontaneously shifted between layers\cite{min2007} and that these instabilities are stronger in systems with larger $J$.  Figure \ref{fig:vortex} shows in-plane projected pseudospin orientation for $J=1,2,3,4$ C2DESs, which correspond to $N=1,2,3,4$ ABC-stacked graphene multilayers. Note that for $J>1$, the arrows in the core of the momentum space vortex have rotated spontaneously toward $\hat{z}$ or $-\hat{z}$ direction indicating the spontaneous charge transfer between layers. The present work identifies ABC stacked multilayer graphene as the most likely candidate for this particular type of exotic broken symmetry state.  
Other types of broken symmetry might occur for other stacking sequences, especially in  the quantum Hall regime. 

\section{Conclusions}

We have shown that $N$-layer graphene at intermediate and strong magnetic fields has a strong tendency towards the appearance of $N$ spin and orbitally degenerate Landau levels very close to $E=0$.  
This property should lead to strong quantum Hall effects at $\pm (4e^2/h) N/2$ in many $N$-layer stacks. The origin of this behavior is the following {\em chirality sum rule}: 
i) The low-energy bands of multilayer graphene can be decomposed into $N_{D}$ doublets with chirality $J_{i}$.  
ii) Although $N_{D}$ depends on the stacking sequence,
$\sum_{i=1}^{N_{D}} \, J_{i} = N$ 
in an $N$-layer stack.  

The chirality sum rule applies precisely only to idealized 
models with only nearest-neighbor intralayer and interlayer tunneling.  It nevertheless 
suggests the likelihood of interesting interaction physics and broken symmetry ground states 
in many neutral or weakly doped multilayer graphene samples.   

\section*{Acknowledgements}
This work was supported by NSF-NRI SWAN and the Welch Foundation.


\begin{thebibliography}{99}

\bibitem{geim2007a}
A. K. Geim and K. S. Novoselov, Nature Materials {\bf 6} (2007), 183.

\bibitem{geim2007b}
A. K. Geim and A. H. MacDonald, Phys. Today {\bf 60} (8) (2007), 35.

\bibitem{novoselov2004}
K. S. Novoselov, A. K. Geim, S. V. Morozov, D. Jiang,
Y. Zhang, S. V. Dubonos, I. V. Grigorieva, and A. A. Firsov,
Science {\bf 306} (2004), 666.

\bibitem{berger2004}
C. Berger, Z. Song, T. Li, X. Li, A. Y. Ogbazghi, R. Feng,
Z. Dai, A. N. Marchenkov, E. H. Conrad, P. N. First and W. A. de
Heer, Phys. Chem. B {\bf 108} (2004), 19912.

\bibitem{ohta2006}
T. Ohta, A. Bostwick, T. Seyller, K. Horn, and E. Rotenberg, 
Science {\bf 313} (2006), 951.

\bibitem{rycerz2007}
A. Rycerz, J. Tworzydl and C. W. J. Beenakker, Nature Phys. {\bf 3} (2007), 172.

\bibitem{cheianov2007}
V. V. Cheianov, V. Fal'ko and B. L. Altshuler, Science {\bf 315} (2007), 1252.

\bibitem{novoselov2005}
K. S. Novoselov, A. K. Geim, S.V. Morozov, D. Jiang,
M. I. Katsnelson, I. V. Grigorieva, S. V. Dubonos and
A. A. Firsov, Nature {\bf 438} (2005), 197.

\bibitem{zhang2005}
Y. Zhang, Y. W. Tan, H. L. Stormer and P. Kim, Nature {\bf 438} (2005), 201.

\bibitem{min2008} 
The present article is an expanded version of Hongki Min and 
A. H. MacDonald, Phys. Rev. B {\bf 77} (2008), 155416.

\bibitem{hass2008}
J. Hass, F. Varchon, J. E. Mill\H{a}n-Otoya, M. Sprinkle,
N. Sharma, W. A. de Heer, C. Berger, P. N. First, L. Magaud and
E. H. Conrad, Phys. Rev. Lett. {\bf 100} (2008), 125504.

\bibitem{charlier1992}
J. C. Charlier, J. P. Michenaud and X. Gonze, Phys. Rev. B {\bf 46} (1992), 4531.

\bibitem{ritger1968}
P. D. Ritger and N. J. Rose, \textit{Equations with
Applications} (McGraw-Hill Book Company, New York, 1968).

\bibitem{guinea2006} 
F. Guinea, A. H. Castro Neto and N. M. R. Peres, Phys. Rev. B {\bf 73} (2006), 245426.

\bibitem{mccann2006}
E. McCann and V. I. Fal'ko, 
Phys. Rev. Lett. {\bf 96} (2006), 086805.

\bibitem{sakurai1994}
J. J. Sakurai, \textit{Modern Quantum Mechanics} (Addison Wesley, Reading, 1994).

\bibitem{koshino2007} 
M. Koshino and T. Ando, Phys. Rev. B {\bf 76} (2007), 085425.

\bibitem{manes2007} 
J. L. Ma\~{n}es, F. Guinea and M. A. H. Vozmediano, Phys. Rev. B {\bf 75} (2007), 155424.

\bibitem{nakamura2008} 
M. Nakamura and L. Hirasawa, Phys. Rev. B {\bf 77} (2008), 045429.

\bibitem{abanin2006} 
D. A. Abanin, P. A. Lee and L. S. Levitov, Phys. Rev. Lett. {\bf 96} (2006), 176803.

\bibitem{nomura2006} 
K. Nomura and A. H. MacDonald, Phys. Rev. Lett. {\bf 96} (2006), 256602.

\bibitem{alicea2006} 
J. Alicea and M. P. A. Fisher, Phys. Rev. B {\bf 74} (2006), 075422.

\bibitem{goerbig2006} 
M. O. Goerbig, R. Moessner and B. Doucot, Phys. Rev. B {\bf 74} (2006), 161407.

\bibitem{yang2006} 
K. Yang, S. Das Sarma and A. H. MacDonald, Phys. Rev. B {\bf 74} (2006), 075423.

\bibitem{barlas2008} 
Y. Barlas, R. Cote, K. Nomura and A. H. MacDonald, Phys. Rev. Lett. {\bf 101} (2008), 097601.

\bibitem{min2007} 
Hongki Min, G. Borghi, M. Polini and A. H. MacDonald, Phys. Rev. B {\bf 77} (2008), 041407.

\end{thebibliography}
\end{document}